\documentclass[12pt]{article}
\usepackage{graphicx}
\usepackage{cite}
\newcommand{\dissum}[2]{\displaystyle \sum_{#1}^{#2}}

\newcommand{\fnd}[2]{\frac{\textstyle #1}{\textstyle #2}}
\newcommand{\xrm}[1]{{\textstyle \mbox{\rm #1}}}
\newcommand{\bm}[1]{\mbox{\boldmath $#1$}}
\newcommand{\abs}[1]{\left| #1\right|}
\newcommand{\ket}[1]{\mbox{$\left| #1\right\rangle$}}
\newcommand{\bracket}[2]{\mbox{$\left\langle #1\left| #2\right.\right
\rangle$}}
\newcommand{\braket}[3]{\mbox{$\left\langle #1\left|
#2\right| #3\right\rangle$}}
\newcommand{\bra}[1]{\mbox{$\left\langle #1\right|$}}
\newcommand{\x}[1]{{\textstyle #1}}
\newcommand{\Real}[1]{\Re {\it e}(#1 )}
\newcommand{\Imag}[1]{\Im {\it m}(#1 )}

\begin{document} \baselineskip .7cm
\title{Reconciling the Light Scalar Mesons \\ with Breit-Wigner Resonances
as well as the Quark Model}
\author{
Eef van Beveren\\
{\normalsize\it Centro de F\'{\i}sica Te\'{o}rica}\\
{\normalsize\it Departamento de F\'{\i}sica, Universidade de Coimbra}\\
{\normalsize\it P-3000 Coimbra, Portugal}\\
{\small eef@teor.fis.uc.pt}\\ [.3cm]
\and
George Rupp\\
{\normalsize\it Centro de F\'{\i}sica das Interac\c{c}\~{o}es Fundamentais}\\
{\normalsize\it Instituto Superior T\'{e}cnico, Edif\'{\i}cio Ci\^{e}ncia}\\
{\normalsize\it P-1049-001 Lisboa Codex, Portugal}\\
{\small george@ajax.ist.utl.pt}\\ [.3cm]
{\small PACS number(s): 11.80.Et, 12.40.Yx, 13.75.Lb, 14.40.-n}
}
\maketitle

\begin{abstract}
Resonances appearing in hadronic scattering processes are described by a
two-phase model.
In the one phase, scattering products are observed, whereas the other phase
describes confinement. A so-called ``Resonance-Spectrum Expansion'' is
derived, containing expressions that resemble Breit-Wigner formulae.
This method also provides a straightforward explanation for the origin of the
light scalar mesons without requiring extra degrees of freedom.
\end{abstract}

\section{Introduction}

For more than three decades now, the light scalar mesons have been puzzling
both experimentalists
\cite{Montanet2003,NUCLEX0302007,HEPPH0301126,HEPPH0302137}
and theorists
\cite{HEPPH0210431,PLB559p49,HEPPH0212361,HEPPH0212117,HEPPH0302059,HEPPH0303248,HEPPH0304031,HEPPH0303223}.
On the experimental side, one is faced
with a highly disparate set of a few well-established, relatively narrow
resonances, as well as some very broad, seemingly non-Breit-Wigner-like
structures in $S$-wave meson-meson scattering. On the other hand, theory
appears to have tremendous difficulties in reproducing these states on the
basis of some microscopic quark substructure. For instance, the naive quark
model would describe these states as confined $P$-wave $q\bar{q}$ systems,
thus putting the masses of the lightest nonet at about 1.3 GeV upwards, and not
below 1 GeV as found experimentally.

The first consistent attempt to describe the lowest scalars as a usual meson
nonet was undertaken by R.~L.~Jaffe in 1977, in the framework of the MIT bag
model \cite{PRD15p267}. In this approach, the lightest exotic $q^2\bar{q}^2$
scalar states indeed form a nonet, and are therefore also called
crypto-exotics. Moreover, due to a very large, attractive color-magnetic
interaction term, the central mass values of the corresponding bag states are
shifted downwards several hundreds of MeV, thus being in rough agreement with
the real parts of the light scalar mesons listed in the PDG tables
\cite{PRD66p010001}, including the $f_0(600)$ or $\sigma$ meson, in those days
called the $\varepsilon$. However, it is not at all clear how to couple these
stable multiquark states to the physical thresholds, and especially what
influence this would have not only on the widths but also on the real part of
the spectrum. Then, in 1982, M.~D.~Scadron presented the first theoretical
description of the light scalars as a nonet of $q\bar{q}$ states
\cite{PRD26p239}. In this
work, the mechanism for producing light scalar mesons is the spontaneous
breaking of chiral symmetry, at the same time responsible for the vanishing of
the pion mass in the chiral limit. However, with the advent
of chiral perturbation theory (ChPT), the $\sigma$ and $\kappa$ mesons
fell out of favor, and for many years it was considered unnecessary, perhaps
even undesirable, to have a complete light scalar nonet. Only with the
undeniable mounting of experimental evidence, first for the $f_0(600)$
\cite{NPB320p1} and recently also for the $K_0^*(800)$ ($\kappa$)
\cite{HEPEX0204018}, many model builders started to rehabilitate the light
scalar nonet, even some ChPT practitioners \cite{PRD59p074001}. This
development also led to a revival of interest in our 1986 model prediction of
the complete light nonet, in a unitarized Schr\"{o}dinger formalism
\cite{ZPC30p615}.

In modern approaches towards fundamental interactions and scattering of
elementary particles, the Schr\"{o}dinger equation has become obsolete.
However, as we may learn from the long-standing difficulties involving the
light scalar mesons, an analysis based on the Schr\"{o}dinger equation, like
the one employed in  Refs.~\cite{ZPC30p615,EPJC22p493} can be very clarifying.
Unfortunately, it is then necessary to enter into the details of solving the
corresponding set of coupled second-order differential equations, a technique
which has become obsolete as well.

\begin{table}[ht]
\begin{center}
\begin{tabular}{|c|c||l|l|l|}
\hline\hline & & & & \\ [-0.3cm]
n & mass (MeV) & $I=1$ & $I=1/2$ & $I=0$ \\
\hline & & & & \\ [-0.3cm]
0 & 1.4 & $a_{0}$(1450) \cite{PRD66p010001} &
$K^{\ast}_{0}$(1430) \cite{PRD66p010001} &
$f_{0}$(1370) \cite{PRD66p010001} $f_{0}$(1500) \cite{PRD66p010001} \\
& & $a_{1}$(1260) \cite{PRD66p010001} & &
$f_{1}$(1285) \cite{PRD66p010001} $f_{1}$(1420) \cite{PRD66p010001} \\
& & $a_{2}$(1320) \cite{PRD66p010001} &
$K^{\ast}_{2}$(1430) \cite{PRD66p010001} &
$f_{2}$(1270) \cite{PRD66p010001} $f_{2}$(1430) \cite{PRD66p010001} \\
\hline & & & & \\ [-0.3cm]
1 & 1.75 & & $K^{\ast}_{0}$(1950) \cite{PRD66p010001} &
$f_{0}$(1710) \cite{PRD66p010001} $f_{0}$(1770) \cite{PLB449p154} \\
& & $a_{1}$(1640) \cite{PRD66p010001} & &
$f_{1}$(1510) \cite{PRD66p010001} \\
& & $a_{2}$(1700) \cite{PRD66p010001} &
$K^{\ast}_{2}$(1980) \cite{PRD66p010001} &
$f_{2}'$(1525) \cite{PRD66p010001}
$f_{2}$(1565) \cite{PRD66p010001} \\
& & & & $f_{2}$(1640) \cite{PRD66p010001}
$f_{2}$(1810) \cite{PRD66p010001}\\
\hline & & & & \\ [-0.3cm]
2 & 2.1 & $a_{0}$(2025) \cite{PLB517p261} & &
$f_{0}$(2020) \cite{PRD66p010001}
$f_{0}$(2200) \cite{PRD66p010001} \\
& & & & \\
& & & & $f_{2}$(1910) \cite{PRD66p010001} $f_{2}$(1950) \cite{PRD66p010001} \\
& & & & $f_{2}$(2010) \cite{PRD66p010001} $f_{2}$(2150) \cite{PRD66p010001} \\
\hline & & & & \\ [-0.3cm]
3 & 2.45 & & & $f_{0}$(2330) \cite{PRD66p010001} \\
& & & & \\
& & & &
$f_{2}$(2300) \cite{PRD66p010001} $f_{2}$(2340) \cite{PRD66p010001} \\
\hline\hline
\end{tabular}
\end{center}
\caption[]{\small The classification of the $J^{P}=(0,1,2)^{+}$
resonances as a function of radial excitation $n$.
The ground states have $n=0$.}
\label{f0}
\end{table}

In Table~\ref{f0}, we give a simple classification for the positive-parity
mesons, based on the model we describe in Sec.~\ref{coupledchannels}.
Each resonance is supposed to originate from a pure confined $q\bar{q}$
state, with quantum numbers $s=1$ for the total intrinsic $q\bar{q}$ spin,
and $\ell =1$ or 3 for the relative orbital angular momentum of the
$q\bar{q}$ pair. In Sec.~\ref{coupledchannels} we explain what we exactly mean
by \em ``originating from'', \em and how this can be consistently described
within a model for meson-meson scattering.
By just varying one model parameter, i.e., the coupling $\lambda$, letting it
decrease from its ``physical'' value $\lambda_{ph}$ towards zero,
the model's spectrum turns from the experimentally observed resonances
into a genuine $q\bar{q}$ confinement spectrum.

For the radial confinement spectrum we have chosen equal spacings in
Table~\ref{f0}, with level splittings of 350 MeV and a ground state ($n=0$) at
1.4 GeV.
The $\ell =3$ radial excitations start out at 1.75 GeV.
Such a spectrum corresponds to the one of a harmonically oscillating
quark-antiquark pair. Harmonic confinement is not essential to our model, but
the cross sections, phase shifts and electromagnetic transition rates resulting
from the full model are in reasonable agreement with experiment
\cite{ZPC30p615,PRD27p1527,PRD44p2803}.

Non-strange ($n\bar{n}$) and strange ($s\bar{s}$) configurations
double the number of isoscalar states into $SU(3)$-flavor singlets
and octets. But one should be be aware that all states get mixed through the
meson loops in our model (see Sec.~\ref{QExchandMLoops}).
Hence, like in Nature we will not obtain pure angular, radial, or flavor
excitations in our model calculations.

In Table~\ref{radex} we have schematically indicated how many
states one must expect at each mass level, except for the ground
states where $\ell =3$ excitations are absent.

\begin{table}[ht]
\begin{center}
\begin{tabular}{|l||l|l|l|}
\hline\hline & & & \\ [-0.3cm]
& $I=1$ & $I=1/2$ & $I=0$ (singlet, octet)\\
\hline & & & \\ [-0.3cm]
& $a_{0}$ & $K^{\ast}_{0}$ &
$f_{0}{\{ 1\}}$ $f_{0}{\{ 8\}}$ \\
$\ell =1$ & $a_{1}$ & $K^{\ast}_{1}$ &
$f_{1}{\{ 1\}}$ $f_{1}{\{ 8\}}$ \\
& $a_{2}$ & $K^{\ast}_{2}$ &
$f_{2}{\{ 1\}}$ $f_{2}{\{ 8\}}$ \\
& & & \\ [-0.3cm]
\hline & & & \\ [-0.3cm]
$\ell =3$ & $a_{2}$ & $K^{\ast}_{2}$ &
$f_{2}{\{ 1\}}$ $f_{2}{\{ 8\}}$ \\ [0.2cm] \hline\hline
\end{tabular}
\end{center}
\caption[]{\small The expected number of $J^{P}=(0,1,2)^{+}$
meson resonances at each level of radial excitation.}
\label{radex}
\end{table}

Meson loops influence, moreover, the precise resonance shapes.
Some come out broad, others narrower,
Also, the central resonance positions may shift substantially (100--300 MeV)
with respect to the underlying $q\bar{q}$ confinement spectrum.

From Tables~\ref{f0} and \ref{radex} we may conclude that
the observed positive-parity mesonic resonances can easily be accommodated
in a quark model, contrary to what has been claimed in recent literature
\cite{PLB541p22,HEPPH0301012}.
Notice in particular the claim in Ref.~\cite{PLB541p22} that \em
``far too many $0^{++}$ resonances were established,
to be accommodated in the ground-state scalar nonet'', \em
a few lines further on followed by the remark \em ``\ldots a missing state''.
\em Inspection of Table~\ref{f0} reveals that still many states are missing,
especially in the $I=1$ and $I=1/2$ sectors.
However, thanks to glueball searches we nowadays have a much better knowledge
of the $I=0$ sector than twenty years ago.
A classification of the mesonic resonances in this sector can
be satisfactorily achieved assuming quark degrees of freedom.
That does not necessarily imply the absence of other configurations.
From lattice calculations it is becoming clear that it may even be
very hard to disentangle the various configurations existing in $f_{0}$
systems \cite{HEPLAT0210012}.
Hence, the $f_{0}$s could, in principle, be mixtures of $q\bar{q}$,
glueballs \cite{HEPPH0302133},
$\left( q\bar{q}\right)^{2}$, meson-meson states, hybrids, etc.,
moreover in all possible color configurations.
Nevertheless, for the \em classification \em \/of positive-parity mesonic
resonances we only need quark degrees of freedom.

However, our table does not contain the light scalar mesons, that is, the
$f_0$(600) ($\sigma$ meson), the $f_0$(980) and $a_0$(980), and the recently
confirmed \cite{HEPEX0204018} $K_0^\ast$(800) ($\kappa$ meson). But we have
shown that such a complete nonet below 1 GeV is inevitable in the quark model,
due to the $^3P_0$ mechanism \cite{ZPC30p615,EPJC22p493}, besides the usual
nonet in the mass region 1.3--1.5 GeV, consisting of the $f_{0}$(1370),
$a_{0}$(1450), $K_{0}^{\ast}$(1430), and $f_{0}$(1500). Notwithstanding
the dissent about the correct interpretation of the individual states,
these two scalar nonets are the only ones that seem to be complete from the
experimental point of view.
Several members of the other nonets still have to be found.
Table~\ref{f0} shows in which energy region we expect the missing
$a_{0}$s and $K_{0}^{\ast}$s to be observed in experiment.

Let us now dwell somewhat more upon the understanding of the scalar mesons,
and compare their situation with positronium. Imagine a teacher asking his/her
students, after having added a fictitious ground state, to invent a theory
that explains the positronium spectrum.
Probably only the cleverest students will discover the malice of their
teacher. The others will invent whatever model it takes to get an explanation
for the false positronium spectrum. In the scalar-meson spectrum, at least two
such extra states seem to exist, namely the firmly established, relatively
narrow $f_{0}(980)$ and $a_{0}(980)$ resonances, which are therefore also the
most controversial states for theorists. As an illustration of the general
confusion here, let us just mention e.g.\ the ``$K\bar{K}$-molecule'' approach
of Ref.~\cite{PRL77p2332}, producing both the $f_0(980)$ and $a_0(980)$ as
dynamical meson-meson resonances due to \em strong $t$-channel
attraction \em \/in the $K\bar{K}$ system, the relativistic quark model of
Ref.~\cite{PLB361p160}, in which only the $f_0(980)$ is described as a
$q\bar{q}$ state, owing its low mass to a strong \em instanton-induced \em
\/interaction, whereas the $a_0(980)$ is supposed to be a $K\bar{K}$ molecule,
and finally the confining NJL model of Refs.~\cite{PRD63p014019,PRD65p114011},
which obtains
the $f_0(980)$ as well as the $a_0(980)$ as light $q\bar{q}$ scalars, thanks to
an attractive \em 't Hooft \em \/interaction (see, however,
Ref.~\cite{PRD65p078501}).  Without these two resonances,
the scalar meson spectrum would seem to start off with a ground state at about
1.3 GeV and could then be simply explained by the naive quark model, provided
one assumed the very broad $\sigma(600)$ and $\kappa(800)$ resonances to be
of whatever dynamical, but not $q\bar{q}$ origin. On the other hand, if one
insists on taking the light scalars as the ground states of the spectrum, then,
at first sight, it seems very hard to achieve a unified description of all
scalar mesons, not to speak of including pseudoscalar, vector, and other
mesons, too.

The main purpose of the present paper is to demonstrate that it is really
possible to uniformly describe the whole nonet of light scalar mesons,
including the broad $\sigma$ and $\kappa$ structures, in a $^3P_0$-modified
Breit-Wigner-like framework, on the basis of $q\bar{q}$ and meson-meson degrees
of freedom only.  However, these states
will turn out to be \em not \em \/the naive ground states of the scalar-meson
spectrum \cite{ZPC30p615,EPJC22p493,EPJC10p469}. The formalism we shall use
is just the old-fashioned Schr\"{o}dinger equation, often considered
``unworthy'' in hadronic physics nowadays, but which nonetheless is perfectly
suited to describe phenomena like (virtual) bound states, resonances, threshold
behavior, Riemann sheets, and so forth. Of course, we do not propose this
technique so as to substitute the modern nonperturbative methods for handling
QCD. It is just intended to pinpoint the structure of the scalar-meson
spectrum, and to show how powerful a simple, intuitive approach can be, as long
as one includes the relevant degrees of freedom. For that purpose, one
unavoidably has to go through some widely forgotten calculus, in order
to obtain an analytic expression for the $S$ matrix from a set of
coupled-channel equations. Therefore, we shall choose the simplest possible
interactions that still contain the essence of the physics, without eclipsing
the latter by a heap of opaque equations.

The organization of this paper is as follows. In Sec.~2 we formally present
the coupled-channel equations linking the confined $q\bar{q}$ sector to the
free two-meson sector. Section 3, in combination with Appendices A, B, and C,
is devoted to a detailed analytical derivation of the $S$ matrix and the
scattering phase shift for a two-channel model, with arbitrary confinement and
a delta-shell interaction to mimic the $^3P_0$ transitions between the
$q\bar{q}$ and meson-meson channels. The here derived formula, which we call
``Resonance-Spectrum Expansion'', is central to the ensuing analysis in
the remainder of the paper. In Sec.~4 we discuss how to extract
results for bound states, resonances, and scattering observables from the
$S$ matrix for the full, multichannel model. In Sec.~5 we describe the general
connection between the discrete confinement spectrum and the $S$-matrix poles
corresponding to physical bound states and resonances, depending on the
threshold energies. Section~6 explains the difference in threshold behavior
of $S$-wave and $P$-wave poles. In Sec.~7 we present a detailed discussion
of the behavior of $S$-matrix poles, as a function of the relative momentum
$k$ or the energy $E$, for the light scalar mesons, in particular the
$K_0^*(800)$ and the $a_0(980)$. Our conclusions on the nature of the scalar
mesons are drawn in Sec.~8.

\section{Coupled channels}
\label{coupledchannels}

Let us consider a mesonic system which may appear in two different phases,
and a mechanism allowing transitions from one phase to the other.
In one phase the system consists of two noninteracting mesons, in the other
phase of two permanently bound particles representing a
quark-antiquark system.
We will refer to the former phase as {\it free}, to the latter as
{\it confined}.
The communication between the two  phases we describe through a short-range
potential $V_{t}$.
In the interaction region, which has the dimension of about 1 fm, we allow
both phases to coexist.
Hence, the wave function of such a system consists of two components there,
$\psi_{f}$ and $\psi_{c}$.
Outside the interaction region the confined component $\psi_{c}$ must, of
course, vanish rapidly.
This can be achieved by a potential $V_{c}$, rising to infinity
with distance, like the linear and the harmonic-oscillator potentials.

Let the Hamiltonians $H_{c}$ and $H_{f}$ describe the dynamics of
the phases of two permanently confined particles and of
two free particles, respectively. Then the following set of coupled
Schr\"{o}dinger equations describes the dynamics of the full system:

\begin{equation}
\left( E-H_{c}\right)\;\psi_{c}\left(\vec{r}\;\right)\; =\;
V_{t}\;\psi_{f}\left(\vec{r}\;\right)
\;\;\;\xrm{and}\;\;\;
\left( E-H_{f}\right)\;\psi_{f}\left(\vec{r}\;\right)\; =\;
\left[ V_{t}\right]^{T}\;\psi_{c}\left(\vec{r}\;\right)
\;\;\; .
\label{cpldeqn}
\end{equation}

Such a philosophy already underlied an elaborate coupled-channel quark
model \cite{ZPC30p615,PRD27p1527,CPC27p377,PRD21p772}, designed
to simultaneously describe mesonic bound-state spectra, resonances, and
meson-meson scattering. However, in spite of the model's success to reproduce
a host of experimental data with a very limited number of parameters, it is
not very suited for the point we wish to make in the present paper, owing to
the specific model choice of the confining $q\bar{q}$ potential, and
especially the rather complicated matrix expressions needed to obtain
$S$-matrix-related observables.
Thus, we shall use here an arbitrary confinement potential, and a
very simple transition potential $V_t$.

\section{Quark exchange and meson loops}
\label{QExchandMLoops}

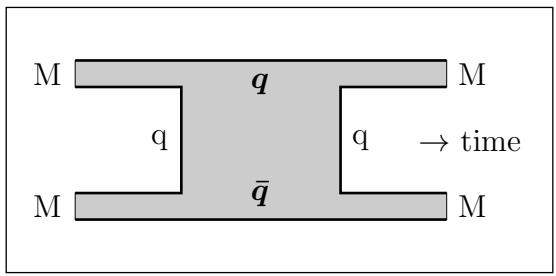
\begin{figure}[ht]
\begin{center}
\begin{picture}(206,100)(-6,0)
\put(15,25){\makebox(0,0)[rc]{M}}
\put(15,75){\makebox(0,0)[rc]{M}}
\put(165,25){\makebox(0,0)[lc]{M}}
\put(165,75){\makebox(0,0)[lc]{M}}
\put(90,25){\makebox(0,0)[bc]{\bm{\bar{q}}}}
\put(90,75){\makebox(0,0)[tc]{\bm{q}}}
\put(55,50){\makebox(0,0)[rc]{q}}
\put(125,50){\makebox(0,0)[lc]{q}}
\put(150,50){\makebox(0,0)[lc]{$\rightarrow$ time}}
\end{picture}
\end{center}
\caption[]{\small The mesons (M) exchange a quark or, equivalently,
a quark-antiquark pair is annihilated,
followed by a second quark exchange, equivalent
to a new quark-antiquark pair being created.}
\label{qexch}
\end{figure}

When, as depicted in Fig~.\ref{qexch}, two interacting mesons exchange
a quark, the resulting system will consist of a valence quark-antiquark pair.
Whether this $q\bar{q}$ pair is going to form a resonance or not will depend
on the quantum numbers of the system and the total available energy.

\noindent
Near a resonance, such a process may be described by scattering phase shifts
of the form

\begin{equation}
\xrm{cotg}\left(\delta_{\ell}(s)\right)\;\approx\;
\fnd{E_{R}-\sqrt{s}}{\Gamma_{R}/2}
\;\;\; ,
\label{cotgdR}
\end{equation}

\noindent
where $E_{R}$ and $\Gamma_{R}$ represent the central
invariant meson-meson mass and the resonance width, respectively.

However, formula (\ref{cotgdR}) is a good approximation for the
scattering cross section only when the resonance shape is not very much
distorted and the width of the resonance is small.
Moreover, the intermediate state in such a process is essentially
a constituent $q\bar{q}$ configuration that belongs to a
confinement spectrum (also referred to as bare or intrinsic states),
and hence may resonate in one of the eigenstates.
This implies that the colliding mesons scatter off
the whole $q\bar{q}$ confinement spectrum of radial, and possibly
also angular excitations, not just off one single state \cite{NC14p951}.
Consequently, a full expression for the phase shifts of
Eq.~(\ref{cotgdR}) should contain all possible eigenstates of such a
spectrum, as long as quantum numbers are respected.
Let us denote the eigenvalues of the relevant part of the spectrum
by $E_{n}$ ($n=0$, $1$, $2$, $\dots$),
and the corresponding eigenstates by $F_{n}$.
Then, following the procedure outlined in Appendices (\ref{Tmtrx}),
(\ref{Delta}), (\ref{Deltaell}), and in Ref.~\cite{EPJC22p493},
we may write for the partial-wave phase shifts the
more general expression

\begin{equation}
\xrm{cotg}\left(\delta (s)\right)\; =\;
\left[ I(s)\;\sum_{n=0}^{\infty}\fnd{
\abs{{\cal F}_{n}}^{2}}{\sqrt{s}-E_{n}}\right]^{-1}\;
\left[ R(s)\;\sum_{n=0}^{\infty}\fnd{
\abs{{\cal F}_{n}}^{2}}{\sqrt{s}-E_{n}}\; -\; 1\right]
\;\;\; ,
\label{cotgdS}
\end{equation}
which we call ``Resonance-Spectrum Expansion''. In $R(s)$ and $I(s)$ we
have absorbed the kinematical factors and the details of two-meson scattering,
and moreover the three-meson vertices.

For an approximate description of a specific resonance, and in
the rather hypothetical case that the three-meson vertices have small
coupling constants, one may single out, from the sum over all
confinement states, one particular state (say number $N$),
the eigenvalue of which is nearest to the
invariant meson-meson mass close to the resonance.
Then, for total invariant meson-meson masses $\sqrt{s}$ in the vicinity
of $E_{N}$, one finds the approximation

\begin{equation}
\xrm{cotg}\left(\delta (s)\right)\;\approx\;
\fnd{\left[ E_{N}\; +\; R(s)\;\abs{{\cal F}_{N}}^{2}\right]\; -\;\sqrt{s}}
{I(s)\;\abs{{\cal F}_{N}}^{2}}
\;\;\; .
\label{cotgdSs}
\end{equation}

\noindent
Formula (\ref{cotgdSs}) is indeed of the Breit-Wigner form~(\ref{cotgdR}),
with the central resonance position and width given by

\begin{equation}
E_{R}\;\approx\; E_{N}\; +\; R(s)\;\abs{{\cal F}_{N}}^{2}
\;\;\;\;\;\xrm{and}\;\;\;\;\;
\Gamma_{R}\;\approx\; 2I(s)\;\abs{{\cal F}_{N}}^{2}
\;\;\; .
\label{ERGR}
\end{equation}

\noindent
In experiment, one observes the influence of the nearest bound state of
the confinement spectrum, as in classical resonating systems.
Nevertheless, Eq.~(\ref{cotgdSs}) is only a good approximation
if the three-meson couplings are small.
Since the coupling of the meson-meson system to quark exchange
is strong, the influence of the higher- and lower-lying excitations is
not negligible.

In the other hypothetical limit, namely of very large couplings, we obtain
for the phase shift the expression

\begin{equation}
\xrm{cotg}\left(\delta (s)\right)\;\approx\;
\fnd{R(s)}{I(s)}
\;\;\; ,
\label{cotgdSl}
\end{equation}

\noindent
which describes scattering off an infinitely hard cavity.

The physical values of the couplings come out somewhere in between
the two limiting cases.
Most resonances and bound states can be classified as stemming from a
specific confinement state \cite{PLB413p137,HEPPH0204328}.
However, some structures in the scattering cross section stem from the
cavity which is formed by quark exchange or pair creation
\cite{EPJC22p493}.
The most notable of such states are the low-lying resonances
observed in $S$-wave pseudoscalar-pseudoscalar scattering
\cite{HEPEX0204018,HEPPH0110156}.

\begin{figure}[ht]
\begin{center}
\begin{picture}(206,100)(-6,0)
\put(15,60){\makebox(0,0)[rc]{M}}
\put(185,60){\makebox(0,0)[lc]{M}}
\put(100,35){\makebox(0,0)[lt]{meson loop}}
\end{picture}
\end{center}
\caption[]{\small The lowest-order self-energy graph for a meson propagator.}
\label{bubble}
\end{figure}
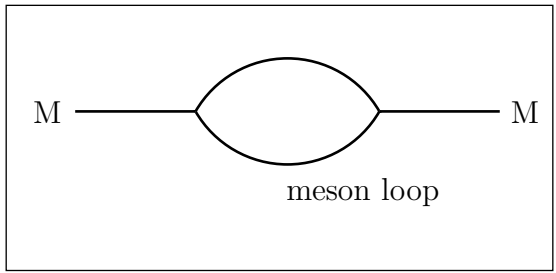

From the discussion above one may conclude
that, to lowest order, the mass of a meson follows from
the quark-antiquark confinement spectrum.
It is, however, well-known that higher-order contributions
to the meson propagator, in particular those from meson loops
as depicted in Fig.~\ref{bubble},
cannot be neglected.

Virtual meson loops give a correction to the meson mass, whereas decay
channels contribute to the strong width of the meson, too.
One obtains for the propagator of a meson the form

\begin{equation}
\Pi(s)\;= \;\fnd{1}
{s\; -\;\left( M_\xrm{confinement}\; +\;\sum\;\Delta M_\xrm{meson loops}
\right)^{2}}
\;\;\; ,
\label{propag}
\end{equation}

\noindent
where $\Delta M$ develops complex values whenever the threshold of a decay
channel gets surpassed.

For the full mass of a meson, all possible meson-meson loops have to be
considered.
A model for meson-meson scattering must therefore include all
possible inelastic channels as well.
Although in principle this could be done, in practice it is not manageable,
unless a scheme exists dealing with all vertices and their relative
intensities (see e.g.\ Ref.~\cite{PRD60p034002} for scalar and
pseudo-scalar meson couplings).
In Refs.~\cite{ZPC21p291,EPJC11p717}
relative couplings have been determined in the
harmonic-oscillator approximation assuming $^{3}P_{0}$ quark exchange.
However, further kinematical factors must be worked out and included.

\section{The spectrum}

The full model consists of an expression for the $K$ matrix
similar to Eq.~(\ref{cotgdS}),
but extended to many meson-meson scattering channels,
several constituent quark-antiquark channels, and more
complicated transition potentials \cite{ZPC30p615,PRD27p1527},
which at the same time
and with the same set of four parameters reproduces bound states,
partial-wave scattering quantities, and the electromagnetic
transitions of $c\bar{c}$ and $b\bar{b}$ systems \cite{PRD44p2803}.

The $K$ matrix can be analytically continued below the various
thresholds, even the lowest one,
with no need to redefine any of the functions involved,
in order to study the singularities of the corresponding scattering matrix.
Below the lowest threshold, these poles show up on the real $\sqrt{s}$
axis, and can be interpreted as the bound states of the coupled system, to be
identified with stable mesons.
For the light flavors one finds this way a nonet of light
pseudoscalars, i.e., the pion, kaon, eta, and eta$'$.
For the heavy flavors, the lowest-lying model poles
can be identified with the $D(1870)$, $D_{s}(1970)$, $\eta_{c}(1S)$,
$J/\psi (1S)$, $\psi (3686)$, $B(5280)$, $B_{s}(5380)$,
$\Upsilon (1S)$, $\Upsilon (2S)$, and $\Upsilon (3S)$.

Above the lowest threshold, the model's partial-wave cross sections
and phase shifts for all included meson-meson channels
can be calculated and compared to experiment, as well as the
inelastic transitions.
Singularities of the scattering matrix come out with negative imaginary
part in the $\sqrt{s}$ plane.
To say it more precisely: out of the many singularities in a rather complex
set of Riemann sheets, some come close enough to the physical real
axis to be noticed in the partial-wave phase shifts and
cross sections.
In fact, each meson-meson channel doubles the number of Riemann
sheets, hence the number of poles.
Consequently, with ten scattering channels one has for each eigenvalue of
the confinement spectrum 1024 poles in 1024 Riemann sheets, out
of which usually only one contains relevant poles in each $\sqrt{s}$
interval in between the thresholds.
Those can be identified with the known resonances,
like the $\rho$ pole in $\pi\pi$ scattering,
or the $K^{\ast}$ pole in $K\pi$ scattering.
However, there may always be a pole in a nearby Riemann sheet
just around the corner of one of the thresholds, which can be
noticed in the partial-wave cross section.
The study of such poles is an interesting subject by itself
\cite{PRD59p074001,NPB587p331}.

Once the four model parameters are adjusted to
the experimental phase shifts and cross sections,
the pole positions can be determined and compared with tables for
meson spectroscopy. For the purpose of the present investigation, we shall
focus next on the singularity structure of the $S$ matrix for the lowest-lying
poles in $S$-wave meson-meson scattering, employing the simplified model of
Sec.~3 above.

\section{Scattering-matrix poles}
\label{scatpoles}

In the hypothetical case of very small couplings for the three-meson
vertices, we obtain poles in the scattering matrix that lie close
to the eigenvalues of the confinement spectrum.
Let us denote by $M_{1}$ and $M_{2}$ the meson masses,
and by $\Delta E$ the difference between the complex-energy pole
of the scattering matrix and the energy eigenvalue $E_{N}$ of
the nearby state of the confinement spectrum.
Using formula (\ref{ERGR}), we obtain

\begin{equation}
\Delta E\;\approx\;
\left\{ R(s)\; -\; iI(s)\right\}\;\abs{{\cal F}_{N}}^{2}
\;\;\; .
\label{DeltaE}
\end{equation}

\noindent
We may distinguish two different cases:
\vspace{0.3cm}

(1) $E_{N}\;>\; M_{1}+M_{2}$ (above threshold),
\vspace{0.1cm}

(2) $E_{N}\;<\; M_{1}+M_{2}$ (below threshold).
\vspace{0.5cm}

\begin{figure}[ht]
\begin{center}
\begin{picture}(400,105)(0,-50)
\put(  0,-50){\line(1,0){400}}
\put(  0, 55){\line(1,0){400}}
\put(  0,-50){\line(0,1){105}}
\put(400,-50){\line(0,1){105}}
\put(0,0){\line(1,0){400}}
\put(100,-2){\line(0,1){4}}
\put(100,-2){\line(1,0){300}}
\put(100, 2){\line(1,0){300}}
\put(380,5){\makebox(0,0)[bl]{\small cut}}
\put(100,5){\makebox(0,0)[bl]{\small threshold}}
\put(100,-5){\makebox(0,0)[tl]{\small $\sqrt{s}=M_{1}+M_{2}$}}
\put(300,0){\makebox(0,0){$\bullet$}}
\put(300,5){\makebox(0,0)[bc]{\small $E_{N}$}}
\put(300,0){\vector(-1,-4){9}}
\put(298,-15){\makebox(0,0)[lt]{\small $\Delta E$}}
\put(290,-40){\makebox(0,0){$\bullet$}}
\put(295,-40){\makebox(0,0)[lc]{\small resonance position}}
\put(5,20){\vector(0,1){20}}
\put(5,43){\makebox(0,0)[cl]{\small $\Imag{\sqrt{s}}$}}
\put(5,20){\vector(1,0){10}}
\put(18,20){\makebox(0,0)[cl]{\small $\Real{\sqrt{s}}$}}
\end{picture}
\end{center}
\caption[]{\small When the confinement state on the real $\sqrt{s}$ axis is in
the scattering continuum, then for small coupling (perturbative regime)
the resonance pole moves into the lower half of the complex $\sqrt{s}$ plane.}
\label{Above}
\end{figure}
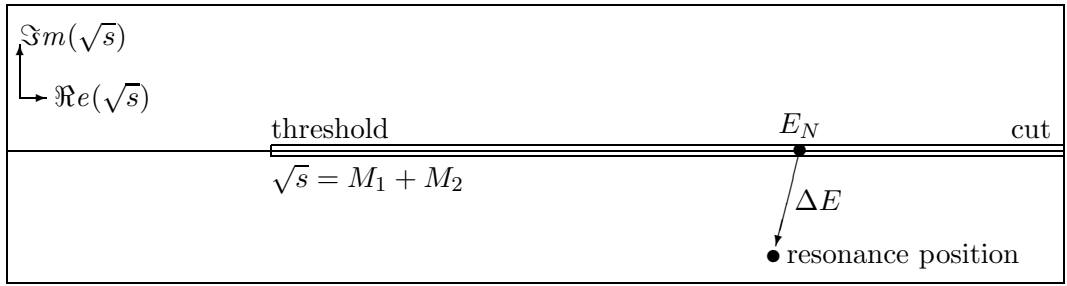

When the nearby state of the confinement spectrum is in the
scattering continuum, then $\Delta E$ has a {\bf negative}
imaginary part and a real part, since both $R(s)$ and $I(s)$
of formula (\ref{DeltaE}) are real, and $I(s)$ is moreover positive.
The resonance singularity of the scattering matrix corresponding to
this situation is depicted in Fig.~\ref{Above}.

Notice that the resonance pole is in the lower half of the
complex-energy plane (second Riemann sheet), as it should be.

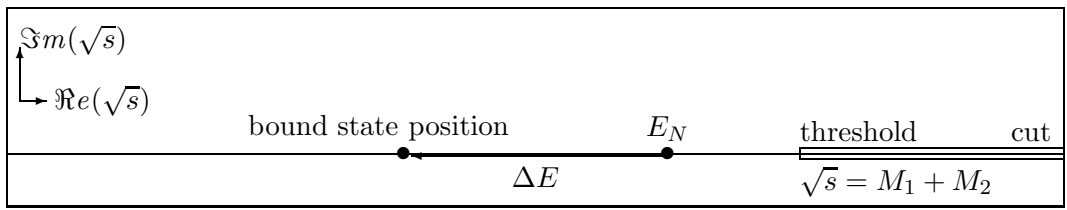
\begin{figure}[ht]
\begin{center}
\begin{picture}(400,75)(0,-20)
\put(  0,-20){\line(1,0){400}}
\put(  0, 55){\line(1,0){400}}
\put(  0,-20){\line(0,1){75}}
\put(400,-20){\line(0,1){75}}
\put(0,0){\line(1,0){400}}
\put(300,-2){\line(0,1){4}}
\put(300,-2){\line(1,0){100}}
\put(300, 2){\line(1,0){100}}
\put(380,5){\makebox(0,0)[bl]{\small cut}}
\put(300,5){\makebox(0,0)[bl]{\small threshold}}
\put(300,-5){\makebox(0,0)[tl]{\small $\sqrt{s}=M_{1}+M_{2}$}}
\put(250,0){\makebox(0,0){$\bullet$}}
\put(250,5){\makebox(0,0)[bc]{\small $E_{N}$}}
\put(250,-1){\vector(-1,0){97}}
\put(200,-5){\makebox(0,0)[tc]{\small $\Delta E$}}
\put(150,0){\makebox(0,0){$\bullet$}}
\put(190,5){\makebox(0,0)[rb]{\small bound state position}}
\put(5,20){\vector(0,1){20}}
\put(5,43){\makebox(0,0)[cl]{\small $\Imag{\sqrt{s}}$}}
\put(5,20){\vector(1,0){10}}
\put(18,20){\makebox(0,0)[cl]{\small $\Real{\sqrt{s}}$}}
\end{picture}
\end{center}
\caption[]{\small When the confinement state on the real $\sqrt{s}$ axis is
below the lowest scattering threshold, then the bound-state singularity
comes out on the real $\sqrt{s}$ axis.}
\label{Below}
\end{figure}

When the nearby state of the confinement spectrum is below
the scattering threshold, then $\Delta E$ has only a real part,
since $I(s)$ turns purely imaginary below threshold, whereas
$R(s)$ remains real.
The bound-state singularity of the scattering matrix corresponding to
this situation is depicted in Fig.~\ref{Below}.

Note that the bound-state pole is on the real axis of
the complex-energy plane, as it should be.

\section{Threshold behavior}

Near the lowest threshold, as a function of the overall coupling
constant, $S$-wave poles behave very differently
from $P$- and higher-wave poles.
This can easily be understood from the effective-range expansion
\cite{PotentialScattering} at the pole position.
There, the cotangent of the phase shift equals $i$.
Hence, for $S$ waves the next-to-lowest-order term in the expansion
equals $ik$ ($k$ represents the linear momentum related to $s$
and to the lowest threshold).
For higher waves, on the other hand, the next-to-lowest-order term in the
effective-range expansion is proportional to $k^{2}$.

\begin{figure}[ht]
\begin{center}
\begin{picture}(320,160)(0,-10)
\put(145,95){\makebox(0,0)[rb]{\small Re($k$)}}
\put(80,145){\makebox(0,0)[lt]{\small Im($k$)}}
\put(31,69){\vector(4,1){10}}
\put(119,69){\vector(-4,1){10}}
\put(69,102){\vector(0,1){10}}
\put(69,43){\vector(0,-1){10}}
\put(64,36){\makebox(0,0)[rt]{\scriptsize virtual}}
\put(64,27){\makebox(0,0)[rt]{\scriptsize bound}}
\put(64,18){\makebox(0,0)[rt]{\scriptsize state}}
\put(64,123){\makebox(0,0)[rt]{\scriptsize bound}}
\put(64,114){\makebox(0,0)[rt]{\scriptsize state}}
\put(120,60){\makebox(0,0)[ct]{\scriptsize resonance}}
\put(120,54){\makebox(0,0)[ct]{\scriptsize pole}}
\put(75,-4){\makebox(0,0)[ct]{$(a)$}}
\put(145,5){\makebox(0,0)[rb]{\small\bf \bm{S} wave}}
\put(315,80){\makebox(0,0)[rb]{\small Re($k$)}}
\put(250,145){\makebox(0,0)[lt]{\small Im($k$)}}
\put(201,64){\vector(4,1){10}}
\put(289,64){\vector(-4,1){10}}
\put(239,102){\vector(0,1){10}}
\put(239,43){\vector(0,-1){10}}
\put(234,36){\makebox(0,0)[rt]{\scriptsize virtual}}
\put(234,27){\makebox(0,0)[rt]{\scriptsize bound}}
\put(234,18){\makebox(0,0)[rt]{\scriptsize state}}
\put(234,123){\makebox(0,0)[rt]{\scriptsize bound}}
\put(234,114){\makebox(0,0)[rt]{\scriptsize state}}
\put(290,60){\makebox(0,0)[ct]{\scriptsize resonance}}
\put(290,54){\makebox(0,0)[ct]{\scriptsize pole}}
\put(245,-4){\makebox(0,0)[ct]{$(b)$}}
\put(315,5){\makebox(0,0)[rb]{\small\bf \bm{P} wave}}
\end{picture}
\end{center}
\caption[]{\small Variation of the positions of scattering-matrix poles
as a function of hypothetical variations in the three-meson-vertex coupling,
for $S$ waves ($a$), and for $P$ and higher waves ($b$).
The arrows indicate increasing coupling constant.}
\label{SPDpoles}
\end{figure}
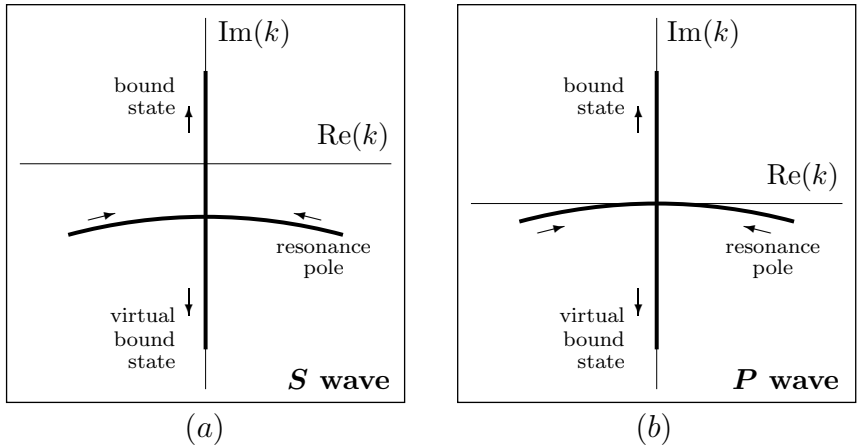

Poles for $P$ and higher waves behave in the complex $k$ plane
as indicated in Fig.~\ref{SPDpoles}$b$.
The two $k$-plane poles meet at threshold ($k=0$).
When the coupling constant of the model is increased, the poles
move along the imaginary $k$ axis.
One pole moves towards negative imaginary $k$, corresponding to
a virtual bound state below threshold on the real $\sqrt{s}$ axis,
but in the wrong Riemann sheet.
The other pole moves towards positive imaginary $k$,
corresponding to a real bound state.

For $S$-wave poles, the behavior is shown in Fig.~\ref{SPDpoles}$a$.
The two $k$-plane poles meet on the negative imaginary $k$ axis.
When the coupling constant of the model is slightly increased,
both poles continue on the negative imaginary $k$ axis,
corresponding to two virtual bound states below threshold on the real
$\sqrt{s}$ axis.
Upon further increasing the coupling constant of the model,
one pole moves towards increasing negative imaginary $k$, thereby
remaining a virtual bound state for all values of the coupling constant.
The other pole moves towards positive imaginary $k$,
eventually passing threshold ($k=0$), thereby turning into a real
bound state of the system of coupled meson-meson scattering channels.
Hence, for a small range of hypothetical values of the coupling constant,
there are two virtual bound states, one of which is very close to
threshold.
Such a pole certainly has noticeable influence on the scattering cross section.

Although we are not aware of any experimental data that could
confirm the above-described threshold behavior of poles,
we suspect this to be possible for atomic transitions in cavities.
Unfortunately, it does not seem likely that in the near future
similar processes can be studied for strong coupling.

\section{The low-lying nonet of \bm{S}-wave poles}

The nonet of low-lying $S$-wave poles behave as described
in Sec.~(\ref{scatpoles}), with respect to variations of the
model's overall coupling constant.
However, they do not stem from the confinement spectrum,
but rather from the cavity.
For small values of the coupling, such poles disappear into the
continuum, i.e., they move towards negative imaginary infinity
\cite{EPJC22p493},
and not towards an eigenstate of the confinement spectrum as in
Fig.~\ref{Above}.

\begin{figure}[ht]
\centerline{\scalebox{0.7}{\includegraphics{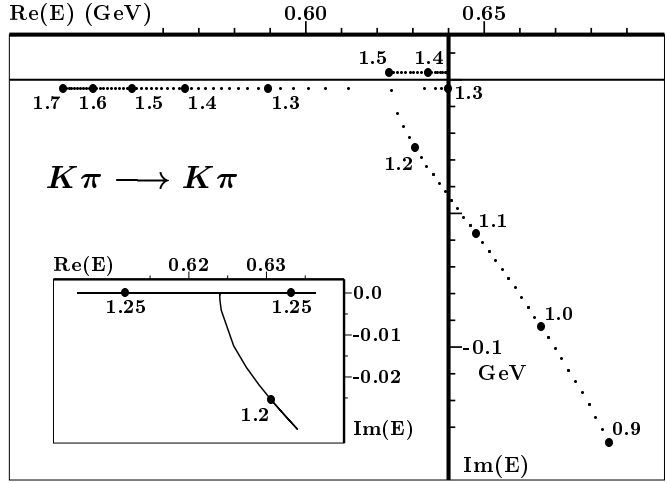}}}
\caption[]{\small Hypothetical movement of the $K_{0}^{\ast}(800)$ pole in
$K\pi$ ($I=1/2$) $S$-wave scattering as a function of the coupling constant
$\lambda$ ($\lambda$ is increased by steps of 0.01 unit, starting from 0.9).
The two branches on the imaginary $k$ axis (see Fig.~\ref{SPDpoles})
both result in poles on the real axis in the $E=\sqrt{s}$ plane.
However, in order to visualize their movement as a function of the
coupling constant, we give virtual bound states ($Im(k)<0$) a small
negative and bound states ($Im(k)>0$) a small positive imaginary part.
In the inset we show in more detail how the poles coming from the
lower half of the complex $E=\sqrt{s}$ plane end up on the real axis when
the coupling constant is increased from 1.2 to 1.25.
}
\label{kappapole}
\end{figure}

In Fig.~\ref{kappapole} we study the hypothetical
movement of the $K_{0}^{\ast}(800)$ pole in $K\pi$
$S$-wave scattering.
The physical value of the coupling constant equals 0.75,
which is not shown in Fig.~\ref{kappapole}.
A figure for smaller values of the coupling constants can be
found in Ref.~\cite{EPJC22p493}.
The physical pole in $K\pi$ isodoublet $S$-wave scattering,
related to experiment \cite{HEPEX0204018},
comes out at $727-i263$ MeV in Ref.~\cite{ZPC30p615}.
Here we concentrate on the threshold behavior of the
hypothetical pole movements in the complex $k$ and $\sqrt{s}$
planes as a function of the coupling constant.
Until they meet on the axis, which is for a value of the
coupling constant slightly larger than 1.24, we have only
depicted the right-hand branch.

The pole corresponding to the one moving downwards
along the imaginary $k$ axis moves to the left on the real
$\sqrt{s}$ axis.
The pole which moves upwards along the imaginary $k$ axis
initially moves towards threshold and then turns back,
following the former pole, but in a different Riemann sheet.
In the inset we clarify the motion of the $S$-matrix singularities
just before and just after they represent virtual and real bound states.
Notice that, since we took 0.14 GeV and 0.50 GeV for
the pion and the kaon mass, respectively, we end up
with a threshold at 0.64 GeV.

It is interesting to note that in a recent work by Boglione and
Pennington \cite{HEPPH0203149} a zero-width state is found
below the $K\pi$ threshold in $S$-wave scattering, instead of the
$K_0^*(800)$ resonance. Here, we would obtain such a state for \em unphysical
\em \/values of the coupling.

\begin{figure}[htbp]
\centerline{\scalebox{0.7}{\includegraphics{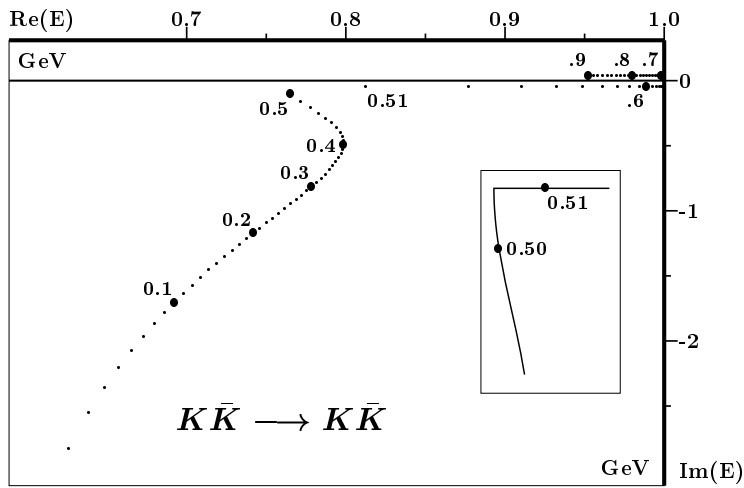}}}
\caption[]{\small Pole movement of the $a_0(980)$ as a function of the coupling
constant $\lambda$ for $K\bar{K}$ ($I=1$) $S$-wave scattering
($\lambda$ is increased by steps of 0.01 unit, starting from 0.02).
As in the case of the $K_{0}^{\ast}(800)$ pole,
the two branches on the imaginary $k$ axis (see Fig.~\ref{SPDpoles})
both result in poles on the real axis in the $E=\sqrt{s}$ plane.
However, in this case we only study the upwards moving branch.
Also here, in order to visualize their movement as a function of the
coupling constant, we give virtual bound states ($Im(k)<0$) a small
negative and bound states ($Im(k)>0$) a small positive imaginary part.
In the inset we show in more detail how the poles coming from the
lower half of the complex $E=\sqrt{s}$ plane end up on the real axis when
the coupling constant is increased from 0.50 to 0.51.
}
\label{a0pole}
\end{figure}

In Fig.~\ref{a0pole} we have depicted the movement of the
$a_ {0}(980)$ pole in $S$ wave $I=1$ $K\bar{K}$ scattering
(threshold at 1.0 GeV) on the upwards-going branch.
One observes a very similar behavior as in the case of $K\pi$
scattering, but with two important differences, to be described next.

The $K_{0}^{\ast}(800)$ poles meet on the real $\sqrt{s}$ axis
only 16 MeV below threshold (see Fig.~\ref{kappapole}),
and for a value of the coupling constant which is well above
the physical value of 0.75, whereas the $a_ {0}(980)$ poles meet
238 MeV below threshold, when the coupling constant
only equals slightly more than 0.5.
At the physical value of the coupling constant, the $a_ {0}(980)$
pole is a real bound state some 9 MeV below threshold.

But there is yet another difference.
Whereas the $K\pi$ channel represents the lowest possible scattering
threshold for the $K_{0}^{\ast}(800)$ system,
$K\bar{K}$ is not the lowest channel for the $a_ {0}(980)$.
In a more complete description, at least all pseudoscalar
meson-meson channels should be taken into account.
One of these is the $\eta\pi$ channel, which has a threshold
well below $K\bar{K}$.
Consequently, upon including the $\eta\pi$ channel in the model, the pole
cannot remain on the real $\sqrt{s}$ axis, but has to acquire
an imaginary part in a similar way as shown in
Fig.~\ref{Above}.
In Ref.~\cite{ZPC30p615} we obtained a resonance-like structure
in the $\eta\pi$ cross section, representing the physical
$a_{0}(980)$. The corresponding pole came out at $962-i28$ MeV.

For the $f_{0}(980)$ system the situation is very similar
to that of the $a_ {0}(980)$.
Assuming a pure $s\bar{s}$ quark content \cite{PLB521p15}, we obtain for the
variation of the corresponding pole in $K\bar{K}$ ($I=0$) $S$-wave scattering
a picture almost equal to the one shown in Fig.~\ref{a0pole}.
However, only in lowest order the $K\bar{K}$ channel can be considered
the lowest threshold for the $f_{0}(980)$ system.
In reality, $s\bar{s}$ also couples to the nonstrange quark-antiquark
isosinglet, namely via the $K\bar{K}$ channel, and hence to $\pi\pi$
\cite{NPB266p451}. Nevertheless, this higher-order coupling turns out to be
rather weak, which implies that the resulting pole does not move far away from
the $K\bar{K}$ bound state.
In Ref.~\cite{ZPC30p615} we obtained a resonance-like structure
in the $\pi\pi$ cross section representing the physical
$f_ {0}(980)$.
The corresponding pole came out at $994-i20$ MeV.

At lower energies, we found for the same cross section a pole
which is the equivalent of the $K_{0}^{\ast}(800)$ system, but now
in $\pi\pi$ isoscalar $S$-wave scattering.
This pole at $470-i208$ MeV can be associated with the
$\sigma$ or $f_0(600)$ meson, since it has the same quantum numbers
and lies in the ballpark of predicted pole positions
in  models for the $\sigma$ (a complete overview of $\sigma$ poles can be
found in Ref.~\cite{HEPPH0201006}).

We do not find any other relevant poles in the energy region
up to 1.0 GeV.

\section{Summary and conclusions}

It should be clear from the foregoing that the light scalar mesons \em do \em
\/allow a description on the basis of normal $q\bar{q}$ states, provided one
accounts for mesonic loops. The crucial point is that, while for the other
mesons the effect of coupling to open and closed meson-meson channels is more
or less perturbative, giving rise to complex or real mass shifts of the
$q\bar{q}$ confinement states, in the case of the scalar mesons the very strong
$^3P_0$ coupling to $S$-wave pseudoscalar-pseudoscalar channels forces
additional, highly non-perturbative poles in the $S$ matrix to approach the
physical region, unhampered by any centrifugal barrier in the two-meson sector.
In other words, these singularities do not stem from the
confinement spectrum, at least not in a simple way, but are rather due to the
$^3P_0$ barrier providing the communication between the $q\bar{q}$ states
and the meson-meson continuum. Nonetheless, these poles give rise to very
pronounced structures in the $S$-wave scattering of pseudoscalar mesons, which
can be either clear-cut resonances ($a_{0}(980)$ and $f_{0}(980)$), or broad
non-resonant bumps ($f_0(600)$ and $K_0^*(800)$).

These conclusions we were able to make more quantitative by employing a
simplified two-channel model with a delta-shell transition potential, allowing
the derivation of a general, closed-form expression of the scattering phase
shift for an \em arbitrary \em \/confinement mechanism. This formula, which we
baptized ``Resonance-Spectrum Expansion'', turns into a standard
Breit-Wigner form in the vicinity of a particular resonance, in the limit of
small coupling and width, whereas it describes hard-sphere scattering in the
limit of large coupling. Application of this formalism to the specific case
of the scalar mesons unmistakably shows that \em all \em \/states of
the light nonet have the same origin, and, in principle, the same threshold
behavior. It just happens that the various thresholds of the different scalars
are very disparate, not in the least due to the small pion mass and hence to
chiral symmetry.

In particular, we found, in the complex-energy plane, a nonet of $S$-matrix
poles representing the $a_{0}(980)$, $f_{0}(980)$, $f_0(600)$, and
$K_0^*(800)$, the latter two having real parts of 0.47 GeV and 0.73 GeV,
respectively, and imaginary parts of 0.21 GeV resp.\ 0.26 GeV.
Whether or not these poles manifest themselves as clear physical resonances
\cite{NPA688p823} is not so relevant here in view of their common nature, as
we have demonstrated in detail for the $K_0^*(800)$ and $a_0(980)$. Besides
these ``non-perturbative'' states, we also found the confinement-ground-state
nonet of scalar mesons $f_{0}(1370)$, $a_{0}(1450)$, $K_{0}^{\ast}(1430)$, and
$f_{0}(1500)$. The latter poles vary as a function of the coupling constant
exactly the way indicated in Fig.~\ref{Above}. For vanishing coupling,
they end up on the real energy axis at the positions of the various
ground-state eigenvalues of the confinement spectrum, which are the
light-flavor $^{3}P_{0}$ states at 1.3 to 1.5 GeV
\cite{PRD61p014015,HEPLAT9805029,NPPS53p236}. Recall that, in contrast, the
poles of the light scalars move towards negative imaginary infinity in this
limit.

In conclusion, we should mention that a coupled-channel or unitarized
approach to the scalar mesons, similar in spirit to our present and previous
\cite{ZPC30p615,EPJC10p469} works, is rapidly gaining advocates
\cite{ZPC68p647,PRL76p1575,PLB462p14,HEPPH0204205}.
Nevertheless, in none of these works a simultaneous discription of the two
lowest scalar nonets is achieved as outlined above, most notably the still
widely contested \cite{ZPC68p647,PRL76p1575,NPA688p823} but now experimentally
confirmed $K_0^*(800)$. As a final remark, let us point out that in our full
model with many coupled channels \cite{ZPC30p615}, all channels contribute to
the states under the resonance, and not just one specific channel. However,
both the full and the simplified model produce very similar masses and widths
for the light scalars when reproducing the experimental phase shifts in the
relevant energy region. This lends additional quantitative support to our
predictions for these observables presented here.
\vspace{0.3cm}

{\bf Acknowledgement}:
We wish to thank F.~Kleefeld for useful discussions on the distribution
of $S$-matrix poles in the complex $E$ and $k$ planes.

This work was partly supported by the
{\it Fun\-da\-\c{c}\~{a}o para a Ci\^{e}ncia e a Tecnologia}
\/of the {\it Minist\'{e}rio da
Ci\^{e}ncia e da Tecnologia} \/of Portugal,
under contract numbers
CERN/\-FIS/\-43697/\-2001
and
CERN/\-FNU/\-49555/\-2002.

\appendix

\section{The \bm{T} matrix for meson-meson scattering}
\label{Tmtrx}

In Eq.~(\ref{cpldeqn}), we must eliminate $\psi_{c}$, since it vanishes at
large distances and is thus {\it unobservable}. Formally, this can be done
in a straightforward way. We then obtain the relation

\begin{equation}
\psi_{f}\left(\vec{r}\;\right)\; =\;
\left( E-H_{f}\right)^{-1}\;\left[ V_{t}\right]^{T}\;
\left( E-H_{c}\right)^{-1}\; V_{t}\;\psi_{f}\left(\vec{r}\;\right)
\;\;\; .
\label{scatteqn}
\end{equation}

By comparison of Eq.~(\ref{scatteqn}) with the usual expressions for
the scattering wave equations, we conclude that the generalized
potential $V$ is here given by

\begin{equation}
V\; =\;\left[ V_{t}\right]^{T}\;\left( E-H_{c}\right)^{-1}\; V_{t}
\;\;\; .
\label{generalpot}
\end{equation}

The matrix elements of the $T$-operator are defined
by the Lippmann-Schwinger equation

\begin{eqnarray}
\lefteqn{T\left({\vec{p}\;},{{\vec{p}\,}'\;}; z\right)\; =\;
V\left({\vec{p}\;},{\vec{p}\,}'\right)\; +}
\nonumber \\ [.3cm] & + &
\int d^{3}k'\;\int d^{3}k\;
V\left({\vec{p}\;},{{\vec{k}\,}'\;}\right)\;
G_{f}\left({{\vec{k}\,}'\;},{\vec{k}\;}; z\right)\;
V\left({\vec{k}\;},{\vec{p}\,}'\right)\; +
\nonumber \\ [.3cm] & + &
\int d^{3}k'''\;\int d^{3}k''\;\int d^{3}k'\;\int d^{3}k\;
V\left({\vec{p}\;},{{\vec{k}\,}'''\;}\right)\;
G_{f}\left({{\vec{k}\,}'''\;},{{\vec{k}\,}''\;}; z\right)\;
V\left({{\vec{k}\,}''\;},{\vec{k}\,}'\right)\;\times
\nonumber \\ [.3cm] & & \;\;\;\times\;
G_{f}\left({{\vec{k}\,}'\;},{\vec{k}\;}; z\right)\;
V\left({\vec{k}\;},{\vec{p}\,}'\right)\; +\;
\dots
\;\;\; ,
\label{Titeration}
\end{eqnarray}

\noindent
where the Green's operator $G_{f}(z)$ corresponds to the self-adjoint free
Hamiltonian $H_{f}$, according to

\begin{equation}
G_{f}\left({{\vec{k}\,}'\;},{\vec{k}\;}; z\right)\; =\;
\braket{{{\vec{k}\,}'\;}}{\left( z-H_{f}\right)^{-1}}{\vec{k}\;}\; =\;
\fnd{2\mu_{f}}{2\mu_{f} z-{k'}^{2}}\bracket{{{\vec{k}\,}'\;}}{\vec{k}\;}
\;\;\; .
\label{Greensfu}
\end{equation}

\noindent
Substitution of relation (\ref{Greensfu}) in expression
(\ref{Titeration}) yields for the $T$-matrix elements the form

\begin{eqnarray}
\lefteqn{T\left({\vec{p}\;},{{\vec{p}\,}'\;}; z\right)\; =\;
V\left({\vec{p}\;},{\vec{p}\,}'\right)\; +\;
\int d^{3}k\; V\left({\vec{p}\;},{\vec{k}\;}\right)\;
\fnd{2\mu_{f}}{2\mu_{f} z-k^{2}}\;
V\left({{\vec{k}\;}},{\vec{p}\,}'\right)\; +}
\label{Titeration1} \\ [.3cm] & + &
\int d^{3}k'\;\int d^{3}k\;
V\left({\vec{p}\;},{{\vec{k}\,}'\;}\right)\;
\fnd{2\mu_{f}}{2\mu_{f} z-{k'}^{2}}\;
V\left({{\vec{k}\,}'\;},{\vec{k}\,}\right)\;
\fnd{2\mu_{f}}{2\mu_{f} z-k^{2}}\;
V\left({\vec{k}\;},{\vec{p}\,}'\right)\; +\;
\dots
\;\;\; .
\nonumber
\end{eqnarray}

At this stage, we must make a choice for the operators $H_{c}$, describing
the confinement dynamics in the interaction region, $H_{f}$, representing the
dynamics of the scattered particles at large distances, and $V_{t}$, which
stands for the transitions between these two sectors. For an arbitrary
spherically symmetric confinement potential $V_{c}$, we define these operators
in configuration space by

\begin{displaymath}
H_{c}\; =\; -\fnd{\nabla^{2}_{r}}{2\mu_{c}}\; +\; m_{q}\; +\; m_{\bar{q}}\;
+\; V_{c}(r)
\;\;\;\;\; ,\;\;\;\;\;
H_{f}\; =\; -\fnd{\nabla^{2}_{r}}{2\mu_{f}}\; +\; M_{1}\; +\; M_{2}
\end{displaymath}

\begin{equation}
\xrm{and}\;\;\;\;\;
V_{t}\; =\; \fnd{\lambda}{2\mu_{c}a}\;\delta\left( r-a\right)
\;\;\; .
\label{Opmodel}
\end{equation}

\noindent
The various mass parameters of Eq.~(\ref{Opmodel}) are defined in
Table~\ref{masses}.
The transition potential $V_{t}$, which provides the communication between
the confined channel and the scattering channel,
is an extreme simplification of potentials \cite{ZPC21p291} that may describe
the breaking of the color string.
Here it is assumed to only act when the particles are at a distance $r=a$,
thus having the form of a sperical delta shell.

\begin{table}[ht]
\begin{center}
\begin{tabular}{|c||l|}
\hline\hline & \\ [-0.3cm]
symbol & definition \\
\hline & \\ [-0.3cm]
$m_{q}$ $\left( m_{\bar{q}}\right)$ & constituent (anti-)quark mass \\
$\mu_{c}$ & reduced mass in confinement channel \\
$M_{1,2}$ & meson masses \\
$\mu_{f}$ & reduced mass in scattering channel \\
\hline\hline
\end{tabular}
\end{center}
\caption[]{\small Definition of mass parameters used in Eq.~(\ref{Opmodel}).}
\label{masses}
\end{table}

In configuration space, we may then write the non-relativistic $2\times 2$
stationary matrix wave equation (\ref{cpldeqn}) in the form

\begin{eqnarray}
\left( -\fnd{\nabla^{2}_{r}}{2\mu_{c}}\; +\; m_{q}\; +\; m_{\bar{q}}\;
+\; V_{c}\; -\; E\right)
\;\psi_{c}\left(\vec{r}\;\right) & = &
-\fnd{\lambda}{2\mu_{c}a}\;
\delta\left( r-a\right)\;\psi_{f}\left(\vec{r}\;\right)
\;\;\; ,\nonumber \\ [.3cm]
\left( -\fnd{\nabla^{2}_{r}}{2\mu_{f}}\; +\; M_{1}\; +\; M_{2}\; -E\right)
\;\psi_{f}\left(\vec{r}\;\right) & = &
-\fnd{\lambda}{2\mu_{c}a}\;
\delta\left( r-a\right)\;\psi_{c}\left(\vec{r}\;\right)
\;\;\; .
\label{c2x2weqCS}
\end{eqnarray}

\subsection{The Born term}

In the momentum representation, Eq.~(\ref{generalpot}) takes the form

\begin{equation}
V\left({\vec{p}\;},{\vec{p}\,}'\right)\; =\;
\bra{\vec{p}\,}\; V\;\ket{{\vec{p}\,}'}\; =\;
\bra{\vec{p}\,}\;\left[ V_{t}\right]^{T}\;
\left( E(p)-H_{c}\right)^{-1}\; V_{t}\;
\ket{{\vec{p}\,}'}
\;\;\; .
\label{MSgeneralpot}
\end{equation}

\noindent
The total center-of-mass energy $E$ and the linear momentum $p$ are, through
Eq.~(\ref{Opmodel}), related by

\begin{equation}
E(p)\; =\;
\fnd{{\vec{p}\;}^{2}}{2\mu_{f}}\; +\; M_{1}\; +\; M_{2}
\;\;\; .
\label{Ep}
\end{equation}

\noindent
We denote the properly normalized eigensolutions of the operator $H_{c}$
(\ref{Opmodel}), corresponding to the energy eigenvalue $E_{n\ell}$, by

\begin{equation}
\bracket{\vec{r}\,}{n\ell m}\; =\;
Y^{(\ell)}_{m}\left(\hat{r}\right)\;
{\cal F}_{n\ell}(r)
\;\;\; ,\;\;\;
\xrm{with $n=0$, $1$, $2$, $\dots$; $\ell =0$, $1$, $2$, $\dots$;
$m=-\ell$, $\dots$, $+\ell$}
\; .
\label{Vccomplete}
\end{equation}

\noindent
So, by letting the self-adjoint operator $H_{c}$ act to the left in
Eq.~(\ref{MSgeneralpot}), we write

\begin{eqnarray}
\bra{\vec{p}\,}\; V\;\ket{{\vec{p}\,}'} & = &
\sum_{n\ell  m}\;\bra{\vec{p}\,}\;
\left[ V_{t}\right]^{T}\;
\ket{n\ell  m}\;\bra{n\ell  m}\;\left( E(p)-H_{c}\right)^{-1}\; V_{t}\;
\ket{{\vec{p}\,}'}
\nonumber\\ [.3cm] & = &
\sum_{n\ell  m}\;\bra{\vec{p}\,}\;\left[ V_{t}\right]^{T}\;
\fnd{\ket{n\ell  m}\;\bra{n\ell  m}}{E(p)-E_{n\ell }}
\; V_{t}\;\ket{{\vec{p}\,}'}
\;\;\; .
\label{MSgeneralpot1}
\end{eqnarray}

\noindent
Next, we insert several times unity to obtain

\begin{eqnarray}
\lefteqn{\bra{\vec{p}\,}\; V\;\ket{{\vec{p}\,}'}\;=\;\sum_{n\ell  m}\;
\int d^{3}r\;\int d^{3}r''\;\int d^{3}r'''\;\int d^{3}r'}
\label{MSgeneralpot2}\\ [.3cm] & &
\times\;\fnd{1}{E(p)-E_{n\ell }}\;
\bracket{\vec{p}\,}{\vec{r}\,}\bra{\vec{r}\,}\;
\left[ V_{t}\right]^{T}\;\ket{{\vec{r}\,}''}
\bracket{{\vec{r}\,}''}{n\ell  m}
\bracket{n\ell  m}{{\vec{r}\,}'''}\bra{{\vec{r}\,}'''}\;
V_{t}\;\ket{{\vec{r}\,}'}\bracket{{\vec{r}\,}'}{{\vec{p}\,}'}
\;\;\; .
\nonumber
\end{eqnarray}

\noindent
Two of the four integrations are trivial, since the local transition
potential has the form

\begin{equation}
\bra{\vec{r}\,}\; V_{t}\;\ket{{\vec{r}\,}'}\; =\;
\fnd{\lambda}{2\mu_{c}a}\;\delta\left( r-a\right)\;
\delta^{(3)}\left(\vec{r}\, -{\vec{r}\,}'\right)
\;\;\; .
\label{Vtnonlocal}
\end{equation}

\noindent
By inserting expression (\ref{Vtnonlocal}) into Eq.~(\ref{MSgeneralpot2}),
also substituting $\bracket{\vec{r}\,}{\vec{p}\,}\; =\;
e^\x{i\vec{p}\cdot\vec{r}}/(2\pi)^{3/2}$,
we get

\begin{eqnarray}
\lefteqn{\bra{\vec{p}\,}\; V\;\ket{{\vec{p}\,}'}\;=\;\sum_{n\ell  m}\;
\int d^{3}r\;\int d^{3}r'}
\label{MSgeneralpot3}\\ [.3cm] & &
\times\;\fnd{1}{E(p)-E_{n\ell }}\;
\fnd{e^\x{-i\vec{p}\cdot\vec{r}}}{(2\pi)^{3/2}}
\fnd{\lambda}{2\mu_{c}a}\;\delta\left( r-a\right)\;
\bracket{\vec{r}\,}{n\ell  m}
\;
\bracket{n\ell  m}{{\vec{r}\,}'}\;
\fnd{\lambda}{2\mu_{c}a}\;\delta\left( r'-a\right)
\fnd{e^\x{i{\vec{p}\,}'\cdot{{\vec{r}\,}'}}}{(2\pi)^{3/2}}
\;\;\; .
\nonumber
\end{eqnarray}

\noindent
Next, we observe that the radial parts of the two remaining integrations
are also trivial, because of the two delta functions. So we twice insert
the expression for the confinement eigenfunctions of Eq.~(\ref{Vccomplete}),
to obtain

\begin{eqnarray}
\lefteqn{\bra{\vec{p}\,}\; V\;\ket{{\vec{p}\,}'}\;=\;
\fnd{1}{(2\pi)^{3}}\;
\left(\fnd{\lambda}{2\mu_{c}a}\right)^{2}\;\sum_{n\ell  m}\;
a^{2}\int d\Omega\; a^{2}\int d\Omega '}
\label{MSgeneralpot4}\\ [.3cm] & &
\times\;\fnd{1}{E(p)-E_{n\ell }}\; e^\x{-i\vec{p}\cdot a\hat{r}}\;
Y^{(\ell )}_{m}\left(\hat{r}\right)\;
{\cal F}_{n\ell }(a)
\;
{Y^{(\ell )}_{m}}^{\ast}\left({\hat{r}\,}'\right)\;
{\cal F}_{n\ell }^{\ast}(a)\;
e^\x{i{\vec{p}\,}'\cdot a{\hat{r}\,}'}
\;\;\; .
\nonumber
\end{eqnarray}

\noindent
For the integrations over the angles we introduce Bauer's formula
\cite{Bauer},

\begin{equation}
e^\x{-i\vec{k}\cdot\vec{r}}\; =\;
\sum_{\lambda ,\mu}4\pi (-i)^{\lambda}j_{\lambda}(kr)
{Y^{(\lambda)}_{\mu}}^{\ast}\left(\hat{r}\right)
Y^{(\lambda)}_{\mu}\left(\hat{k}\right)
\;\;\; ,
\label{Bauerform}
\end{equation}

\noindent
resulting in

\begin{equation}
\int d\Omega\;\left\{
\begin{array}{r}
e^\x{-i\vec{p}\cdot a\hat{r}}\;
Y^{(\ell)}_{m}\left(\hat{r}\right)\\ [.3cm]
e^\x{i{\vec{p}\,}\cdot a{\hat{r}\,}}\;
Y^{(\ell)}_{m}\left({\hat{r}\,}\right)
\end{array}\right\}
\; =\; 4\pi\; j_{\ell}(pa)\;\left\{
\begin{array}{r}
(-i)^{\ell}\; {Y^{(\ell )}_{m}}^{\ast}\left(\hat{p}\right)\\ [.3cm]
(i)^{\ell}\; Y^{(\ell )}_{m}\left({\hat{p}\,}\right)
\end{array}\right\}
\;\;\; .
\label{X01}
\end{equation}

\noindent
Substitution of the relations (\ref{X01}) into Eq.~(\ref{MSgeneralpot4})
leads to the expression

\begin{equation}
\bra{\vec{p}\,}\; V\;\ket{{\vec{p}\,}'}\;=\;
\fnd{\lambda^{2}a^{2}}{2\pi\mu_{c}^{2}}\;
\sum_{n\ell m}\;
\fnd{1}{E(p)-E_{n\ell}}\; {Y^{(\ell)}_{m}}^{\ast}\left(\hat{p}\right)\;
Y^{(\ell)}_{m}\left({\hat{p}\,}'\right)\;
j_{\ell}(pa)\; j_{\ell}(p'a)\;
\abs{{\cal F}_{n\ell}(a)}^{2}
\;\;\; ,
\label{MSgeneralpot5}
\end{equation}

\noindent
where the summation over the magnetic quantum number $m$ can be performed by
the use of the addition theorem, thus shaping the Born term
(\ref{MSgeneralpot}) into its final form

\begin{equation}
\bra{\vec{p}\,}\; V\;\ket{{\vec{p}\,}'}\; =\;
\fnd{\lambda^{2}a^{2}}{8\pi^{2}\mu_{c}^{2}}\;
\sum_{\ell =0}^{\infty}(2\ell +1)\;
P_{\ell}\left(\hat{p}\cdot{\hat{p}\,}'\right)\;
j_{\ell}(pa)\; j_{\ell}(p'a)\;
\sum_{n=0}^{\infty}\;\fnd
{\abs{{\cal F}_{n\ell}(a)}^{2}}
{E(p)-E_{n\ell}}
\;\;\; .
\label{MSgeneralpot6a}
\end{equation}

\subsection{The second-order term}
\label{second}

For the second-order term, we start by substituting the result
(\ref{MSgeneralpot6a}) into the second term of expansion (\ref{Titeration1}),
giving rise to the expression

\begin{eqnarray}
\lefteqn{T^{(2)}\left({\vec{p}\;},{{\vec{p}\,}'\;}; z\right)\; =\;
\int d^{3}k\;
\fnd{\lambda^{2}a^{2}}{8\pi^{2}\mu_{c}^{2}}\;
\sum_{\ell =0}^{\infty}(2\ell +1)\;
P_{\ell}\left(\hat{p}\cdot\hat{k}\,\right)\;
j_{\ell}(pa)\; j_{\ell}(ka)\;
\sum_{n=0}^{\infty}\;\fnd
{\abs{{\cal F}_{n\ell}(a)}^{2}}
{E(p)-E_{n\ell}}}
\nonumber \\ [.3cm] & & \times\;
\fnd{2\mu_{f}}{2\mu_{f} z-k^{2}}\;
\fnd{\lambda^{2}a^{2}}{8\pi^{2}\mu_{c}^{2}}\;
\sum_{\ell '=0}^{\infty}(2\ell '+1)\;
P_{\ell '}\left(\hat{k}\cdot{\hat{p}\,}'\right)\;
j_{\ell '}(ka)\; j_{\ell '}(p'a)\;
\sum_{n'=0}^{\infty}\;\fnd
{\abs{{\cal F}_{n'\ell '}(a)}^{2}}
{E(k)-E_{n'\ell '}}
\nonumber \\ [.3cm] & = &
\left(\fnd{\lambda^{2}a^{2}}{8\pi^{2}\mu_{c}^{2}}\right)^{2}\;
\sum_{\ell =0}^{\infty}(2\ell +1)\;
j_{\ell}(pa)\;
\sum_{n=0}^{\infty}\;\fnd
{\abs{{\cal F}_{n\ell}(a)}^{2}}
{E(p)-E_{n\ell}}\;
\sum_{\ell '=0}^{\infty}(2\ell '+1)\;
j_{\ell '}(p'a)
\nonumber \\ [.3cm] & & \times\;
2\mu_{f}\int d^{3}k\;
P_{\ell}\left(\hat{p}\cdot\hat{k}\,\right)\;
P_{\ell '}\left(\hat{k}\cdot{\hat{p}\,}'\right)\;
\fnd{j_{\ell}(ka)\; j_{\ell '}(ka)}{2\mu_{f} z-k^{2}}\;
\sum_{n'=0}^{\infty}\;\fnd
{\abs{{\cal F}_{n'\ell '}(a)}^{2}}
{E(k)-E_{n'\ell '}}
\;\;\; .
\label{T2def}
\end{eqnarray}

\noindent
The details of the $\vec{k}$ integration are discussed in
Appendix~(\ref{kintegration}).
Since, as required by Eq.~(\ref{calIdef}) below,
$E(k)$ is quadratic in $\vec{k}$,
we find for expression (\ref{T2def}) the result

\begin{eqnarray}
\lefteqn{T^{(2)}\left({\vec{p}\;},{{\vec{p}\,}'\;}\right)\; =\;
\left(\fnd{\lambda^{2}a^{2}}{8\pi^{2}\mu_{c}^{2}}\right)^{2}\;
\sum_{\ell =0}^{\infty}(2\ell +1)\;
j_{\ell}(pa)\;
\sum_{n=0}^{\infty}\;\fnd
{\abs{{\cal F}_{n\ell}(a)}^{2}}
{E(p)-E_{n\ell}}\;
\sum_{\ell '=0}^{\infty}(2\ell '+1)\;
j_{\ell '}(p'a)}
\nonumber \\ [.3cm] & & \times\;
\left( -i\;\fnd{4\pi^{2}\mu_{f}p}{2\ell +1}\right)\;
\delta_{\ell ,\ell '}\;
P_{\ell}\left(\hat{p}\cdot{\hat{p}\,}'\;\right)\;
j_{\ell}(pa)\; h^{(1)}_{\ell}(pa)\;
\sum_{n'=0}^{\infty}\;\fnd
{\abs{{\cal F}_{n'\ell '}(a)}^{2}}
{E(p)-E_{n'\ell '}}
\label{T2result} \\ [.3cm] & = &
-i\;\fnd{\mu_{f}p}{16\pi^{2}}\;
\left(\fnd{\lambda a}{\mu_{c}}\right)^{4}\;
\sum_{\ell =0}^{\infty}(2\ell +1)\;
\; P_{\ell}\left(\hat{p}\cdot{\hat{p}\,}'\;\right)\;
j^{2}_{\ell}(pa)\; h^{(1)}_{\ell}(pa)\; j_{\ell}(p'a)\;
\left[\;
\sum_{n=0}^{\infty}\;\fnd
{\abs{{\cal F}_{n\ell}(a)}^{2}}
{E(p)-E_{n\ell}}\;
\right]^{2}
\;\;\; .
\nonumber
\end{eqnarray}

\subsection{To all orders}
\label{all}

Following steps similar to those in Appendix~(\ref{second}),
it is now straightforward to determine the higher-order contributions
to the expansion (\ref{Titeration1}).
For the full $T$ matrix to all orders, one ends up with the result

\begin{eqnarray}
\lefteqn{T\left({\vec{p}\;},{{\vec{p}\,}'\;}\right)\; =\;}
\label{Tfull} \\ [.3cm] & = &
\fnd{1}{8\pi^{2}}\;
\left(\fnd{\lambda a}{\mu_{c}}\right)^{2}\;
\sum_{\ell =0}^{\infty}(2\ell +1)\;
P_{\ell}\left(\hat{p}\cdot{\hat{p}\,}'\;\right)\;
\fnd
{j_{\ell}(pa)\; j_{\ell}(p'a)\;
\dissum{n=0}{\infty}
\fnd{\abs{{\cal F}_{n\ell}(a)}^{2}}
{E(p)-E_{n\ell}}}
{1+\frac{1}{2}i\mu_{f}p\;
\left(\fnd{\lambda a}{\mu_{c}}\right)^{2}
j_{\ell}(pa)\; h^{(1)}_{\ell}(pa)\;
\dissum{n=0}{\infty}
\fnd{\abs{{\cal F}_{n\ell}(a)}^{2}}
{E(p)-E_{n\ell}}}
\;\;\; .
\nonumber
\end{eqnarray}

\subsection{Scattering matrix and phase shift}
\label{cotdelta}

For radially symmetric interactions, it is useful to define
the partial-wave matrix element $T_{\ell}$ of the on-shell
(${\vec{p}\,}'=\vec{p}$) $T$ matrix, according to
the relation

\begin{equation}
T\left({\vec{p}\;}\right)\; =\;
\sum_{\ell =0}^{\infty}(2\ell +1)\;
P_{\ell}\left(\hat{p}\cdot{\hat{p}\,}'\;\right)\;
T_{\ell}(p)
\;\;\; .
\label{Tpartial}
\end{equation}

\noindent
Hence, also using the result of Eq.~(\ref{Tfull}), we find for the
partial-wave scattering amplitude $S_{\ell}(p)$ the expression

\begin{eqnarray}
S_{\ell}(p) & = &
1-8i\pi^{2}\mu_{f}p\; T_{\ell}(p)
\label{Spartial} \\ [.3cm] & = &
\fnd
{1-\frac{1}{2}i\mu_{f}p\;
\left(\fnd{\lambda a}{\mu_{c}}\right)^{2}
j_{\ell}(pa)\; h^{(2)}_{\ell}(pa)\;
\dissum{n=0}{\infty}
\fnd{\abs{{\cal F}_{n\ell}(a)}^{2}}
{E(p)-E_{n\ell}}}
{1+\frac{1}{2}i\mu_{f}p\;
\left(\fnd{\lambda a}{\mu_{c}}\right)^{2}
j_{\ell}(pa)\; h^{(1)}_{\ell}(pa)\;
\dissum{n=0}{\infty}
\fnd{\abs{{\cal F}_{n\ell}(a)}^{2}}
{E(p)-E_{n\ell}}}
\;\;\; .
\nonumber
\end{eqnarray}

For the partial-wave scattering phase shift $\delta_{\ell}(p)$, defined by
$S_{\ell}(p)\; =\; e^\x{2i\delta_{\ell}(p)}$,
one obtains from Eq.~(\ref{Spartial}) the result

\begin{equation}
\xrm{cotg}\left(\delta_{\ell}(p)\right)\; =\;
\fnd
{\frac{1}{2}\mu_{f}p\;
\left(\fnd{\lambda a}{\mu_{c}}\right)^{2}
j_{\ell}(pa)\; n_{\ell}(pa)\;
\dissum{n=0}{\infty}
\fnd{\abs{{\cal F}_{n\ell}(a)}^{2}}
{E(p)-E_{n\ell}}\; -\; 1}
{\frac{1}{2}\mu_{f}p\;
\left(\fnd{\lambda a}{\mu_{c}}\right)^{2}
j_{\ell}^{2}(pa)\;
\dissum{n=0}{\infty}
\fnd{\abs{{\cal F}_{n\ell}(a)}^{2}}
{E(p)-E_{n\ell}}}
\;\;\; .
\label{partialpshift}
\end{equation}

Formula (\ref{cotgdS}) for the meson-meson scattering phase shift
is based on the latter equation, but not exclusively, as
we discuss in Appendix~(\ref{Deltaell}).

\subsection{Details of the \bm{\vec{k}} integration}
\label{kintegration}

Let us study the momentum-space integration

\begin{eqnarray}
\lefteqn{{\cal I}_{\ell}
\left({\vec{p}\;},{{\vec{p}\,}'\;};\mu\; ;f_{\ell}\right)\; =}
\label{calIdef} \\ [.3cm] & & =
2\mu\;\int d^{3}k\;
P_{\ell}\left(\hat{p}\cdot{\hat{k}\;}\right)\;
P_{\ell '}\left(\hat{k}\cdot{\hat{p}\,}'\right)\;
\fnd{j_{\ell}(ka)\; j_{\ell '}(ka)}{2\mu z-k^{2}}\;
f_{\ell}\left( k^{2}\right)
\nonumber \\ [.3cm] & & =
2\mu\;\int d\Omega_{k}\;
P_{\ell}\left(\hat{p}\cdot{\hat{k}\;}\right)\;
P_{\ell '}\left(\hat{k}\cdot{\hat{p}\,}'\right)\;
\int_{0}^{\infty} k^{2}dk\;
\fnd{j_{\ell}(ka)\; j_{\ell '}(ka)}{2\mu z-k^{2}}
f_{\ell}\left( k^{2}\right)
\;\;\; ,
\nonumber
\end{eqnarray}

\noindent
where $f_{\ell}$ represents an arbitrary well-behaved function of $k^2$.
For the integration over the angles, we can employ the orthogonality of
spherical harmonics.
Hence, we must concentrate on the radial integration, i.e.,

\begin{equation}
\int_{0}^{\infty} k^{2}dk\;
\fnd{j^{2}_{\ell}(ka)}{2\mu z-k^{2}}\;
f_{\ell}\left( k^{2}\right)
\;\;\; .
\label{kmodpart}
\end{equation}

\noindent
We shall show below that the integration can easily be performed, yielding

\begin{equation}
\int_{0}^{\infty} k^{2}dk\;
\fnd{j^{2}_{\ell}(ka)}{2\mu z-k^{2}}\;
f_{\ell}\left( k^{2}\right)\; =\;
\frac{1}{2}\int_{-\infty}^{\infty} k^{2}dk\;
\fnd{j_{\ell}(ka)\; h^{(1)}_{\ell}(ka)}{2\mu z-k^{2}}\;
f_{\ell}\left( k^{2}\right)
\;\;\; ,
\label{kmodpart1}
\end{equation}

\noindent
by using the following properties of the spherical
Bessel and Hankel functions:

\begin{equation}
j_{\ell}\left(e^\x{\pi i}\; ka\right)\; =\;
e^\x{\pi i\ell}\; j_{\ell}(ka)
\;\;\;\xrm{and}\;\;\;\;
h^{(1)}_{\ell}\left(e^\x{\pi i}\; ka\right)\; =\;
e^\x{-\pi i\ell}\; h^{(2)}_{\ell}(ka)
\;\;\; .
\label{BesHanprop}
\end{equation}

\noindent
For large imaginary part of the argument $ka$, the function
$h^{(1)}_{\ell}(ka)$ tends to zero. Therefore, we can close the integration
path in the complex $k$ plane by a non-contributing semicircle in the upper
half plane. If we then set $2\mu z\; =\; (p+i\epsilon )^{2}$,
taking the limit $\epsilon\downarrow 0$ after the integration, the integral
(\ref{kmodpart1}) can be simply computed with Cauchy's residue theorem,
yielding

\begin{eqnarray}
\int_{0}^{\infty} k^{2}dk\;
\fnd{j^{2}_{\ell}(ka)}{2\mu z-k^{2}}\;
f_{\ell}\left( k^{2}\right) & = &
\lim_{\epsilon\downarrow 0}\;
\frac{1}{2}\oint k^{2}dk\;
\fnd{j_{\ell}(ka)\; h^{(1)}_{\ell}(ka)}
{(p+i\epsilon -k)(p+i\epsilon +k)}\;
f_{\ell}\left( k^{2}\right)
\nonumber \\ [.3cm] & = &
-\fnd{i\pi p}{2}\; j_{\ell}(pa)\; h^{(1)}_{\ell}(pa)\;
f_{\ell}\left( p^{2}\right)
\;\;\; .
\label{kmodpart2}
\end{eqnarray}

Putting everything together, we obtain for Eq.~(\ref{calIdef}) the final result

\begin{equation}
{\cal I}_{\ell}
\left({\vec{p}\;},{{\vec{p}\,}'\;};\mu\; ;f_{\ell}\right)\; =\;
-i\;\fnd{4\pi^{2}\mu p}{2\ell +1}\;\delta_{\ell ,\ell '}
\; P_{\ell}\left(\hat{p}\cdot{\hat{p}\,}'\;\right)\;
j_{\ell}(pa)\; h^{(1)}_{\ell}(pa)\;
f_{\ell}\left( p^{2}\right)
\;\;\; .
\label{kintresult}
\end{equation}

\section{The phase shift in the configuration-space approach}
\label{Delta}

The radial wave equation, following from Eq.~(\ref{c2x2weqCS})
by choosing

\begin{equation}
\left(\begin{array}{c} \psi_{c}\left(\vec{r}\;\right) \\ [.3cm]
\psi_{f}\left(\vec{r}\;\right)\end{array}\right)\; =\;
\left(\begin{array}{c}
\fnd{u_{c}(r)}{r}\; Y^{(\ell )}_{\ell_{z}}(\vartheta ,\varphi ) \\ [.3cm]
\fnd{u_{f}(r)}{r}\; Y^{(\ell )}_{\ell_{z}}(\vartheta ,\varphi )
\end{array}\right)
\;\;\; ,
\label{angdeco}
\end{equation}

\noindent
is given by

\begin{equation}
\left(\begin{array}{cc} h_{c} & \lambda V(r)\\ [.3cm]
\lambda V(r) & h_{f}\end{array}\right)\;
\left(\begin{array}{c} u_{c} \\ [.3cm] u_{f}\end{array}\right)\; =\;
E\;\left(\begin{array}{c} u_{c} \\ [.3cm] u_{f}\end{array}\right)
\;\;\; ,
\label{2x2weq2}
\end{equation}

\noindent
where (see Eq.~(\ref{Opmodel}))

\begin{eqnarray}
h_{c} & = & \fnd{1}{2\mu_{c}}
\left( -\fnd{d^{2}}{dr^{2}}\; +\;
\fnd{\ell\left(\ell +1\right)}{r^{2}}\right)\; +\;
m_{q}\; +\; m_{\bar{q}}\; +\; V_{c}(r)
\;\;\; ,
\nonumber \\ [.3cm]
h_{f} & = & \fnd{1}{2\mu_{f}}
\left( -\fnd{d^{2}}{dr^{2}}\; +\;
\fnd{\ell \left(\ell +1\right)}{r^{2}}\right)\; +\;
M_{1}\; +\; M_{2}
\;\;\; , \;\;\;\xrm{and}
\nonumber \\ [.3cm]
V & = & \fnd{1}{2\mu_{c}a}\;\delta\left( r-a\right)
\;\;\; .
\label{2x2f1dlt}
\end{eqnarray}
In this section, we study the solutions of the $2\times 2$ radial wave
equation (\ref{2x2weq2}) in configuration space.

For $r<a$ as well as for $r>a$, one has to solve the two uncoupled
differential equations, equivalent to $\lambda =0$, given by

\begin{equation}
\left(\begin{array}{cc} h_{c} & 0 \\ [.3cm] 0 & h_{f}\end{array}\right)\;
\left(\begin{array}{c} u_{c}\\ [.3cm] u_{f}\end{array}\right)\; =\;
E\;\left(\begin{array}{c} u_{c}\\ [.3cm] u_{f}\end{array}\right)
\;\;\; .
\label{uncoupled}
\end{equation}
At $r=a$ one has the boundary conditions

\begin{equation}
\left\{\begin{array}{l}
\fnd{1}{2\mu_{c}}
\left( -\left.\fnd{d\; u_{c}(r)}{dr}\right|_{r\downarrow a}\; +\;
\left.\fnd{d\; u_{c}(r)}{dr}\right|_{r\uparrow a}\right)\; +\;
\fnd{\lambda}{2\mu_{c}a}u_{f}(a)\; =\; 0 \\ [1cm]
\fnd{\lambda}{2\mu_{c}a}u_{c}(a)\; +\;\fnd{1}{2\mu_{f}}
\left( -\left.\fnd{d\; u_{f}(r)}{dr}\right|_{r\downarrow a}\; +\;
\left.\fnd{d\; u_{f}(r)}{dr}\right|_{r\uparrow a}\right)
\; =\; 0\end{array}\right.
\;\;\; ,
\label{reqa1}
\end{equation}

\noindent
and

\begin{equation}
\left\{\begin{array}{l}
u_{c}(r\uparrow a)\; =\; u_{c}(r\downarrow a)\\ [.3cm]
u_{f}(r\uparrow a)\; =\; u_{f}(r\downarrow a)\end{array}
\right.
\;\;\; .
\label{reqa2}
\end{equation}

\noindent
Further boundary conditions are the usual ones:
the wave functions $u_{c}$ and $u_{f}$ must both vanish at the origin.
Towards infinity, the wave function $u_{c}$ must be damped exponentially, since
$u_{c}$ describes a confined system, whereas, above threshold
($E\;>\; M_{1}+M_{2}$), the wave function $u_{f}$ must
have an oscillating behavior, describing the scattered mesons.

Let us denote by $F_{c,\ell}$ and $G_{c,\ell}$
the solutions of the upper differential
equation of Eq.~(\ref{uncoupled}), which vanish at the origin and fall off
exponentially at infinity, respectively.
The solution $F_{c,\ell}(E,r)$ is regular at the origin,
but is irregular at infinity,
except for some special cases known as the confinement spectrum,
whereas $G_{c,\ell}(E,r)$ behaves regularly at infinity but not at the origin.
For arguments belonging to the confinement spectrum, $F_{c,\ell}$ and
$G_{c,\ell}$ are degenerate and normalizable.

The lower differential equation Eq.~(\ref{uncoupled}) is solved by the
functions $J$ and $N$ defined by

\begin{equation}
J_{\ell}(k,r)\; =\; k^{-\ell}rj_{\ell}(kr)
\;\;\;\xrm{and}\;\;\;
N_{\ell}(k,r)\; =\; k^{\ell +1}rn_{\ell}(kr)
\;\;\; ,
\label{BN}
\end{equation}

\noindent
where the linear momentum $k$ is given by
\begin{equation}
k^{2}\; =\; 2\mu_{f}\left( E-M_{1}-M_{2}\right)
\;\;\; ,
\label{2muEek}
\end{equation}

\noindent
and where $j_{\ell}$ and $n_{\ell}$ are the spherical Bessel and Neumann
functions, respectively.
The solution $J_{\ell}(k,r)$ satisfies the usual boundary condition at the
origin, i.e., $J_{\ell}(r\rightarrow 0)\rightarrow 0$.
For the solutions (\ref{BN}) one has the Wronskian relation

\begin{equation}
W\left( J_{\ell}(k,a),N_{\ell}(k,a)\right)\; =\;
\left[ J_{\ell}(k,r)\fnd{d\; N_{\ell}(k,r)}{dr}-
\fnd{d\; J_{\ell}(k,r)}{dr}N_{\ell}(k,r)\right]_{r\rightarrow a}\; =\; 1
\;\;\; .
\label{WronskianBN}
\end{equation}

A general solution of the Schr\"{o}dinger equation~(\ref{2x2f1dlt}),
which, moreover, satisfies the boundary conditions at the origin and at
infinity, reads

\begin{equation}
\left(\begin{array}{c} u_{c}(E,r)\\ [.3cm] u_{f}(E,r)\end{array}\right)\; =\;
\left\{\begin{array}{lc}
\left(\begin{array}{c} F_{c,\ell}(E,r)\; A_{c}\\ [.3cm]
J_{\ell }\left( k,r\right)\; A_{f}\end{array}\right) & r<a
\\ [1cm]
\left(\begin{array}{c} G_{c,\ell}(E,r)\; B_{c}\\ [.3cm]
\left[
J_{\ell }\left( k,r\right)
k^{2\ell +1}\xrm{cotg}\left(\delta_{\ell }(E)\right)-
N_{\ell }\left( k,r\right)\right]\; B_{f}
\end{array}\right) & r>a
\end{array}\right. \;\;\;\; ,
\label{solution1}
\end{equation}

\noindent
where $A_{c}$, $A_{f}$, $B_{c}$ and $B_{f}$ are normalization constants,
which are not independent because of the boundary conditions~(\ref{reqa1})
and (\ref{reqa2}).
From the boundary conditions~(\ref{reqa2}) we derive the equations

\begin{equation}
\begin{array}{rcl}
F_{c,\ell}(E,a)\; A_{c} & = & G_{c,\ell}(E,a)\; B_{c}
\;\;\; , \\ [.3cm]
J_{\ell }\left( k,a\right)\; A_{f} & = &
\left[
J_{\ell }\left( k,a\right)
k^{2\ell +1}\xrm{cotg}\left(\delta_{\ell }(E)\right)-
N_{\ell }\left( k,a\right)\right]\; B_{f}
\;\;\; .
\label{reqa2a}
\end{array}
\end{equation}

\noindent
Similarly, from the boundary conditions~(\ref{reqa1}) we obtain
\begin{equation}
\begin{array}{rcl}
\fnd{1}{2\mu_{c}}\left( G'_{c}(E,a)B_{c}-F'_{c}(E,a)A_{c}\right)
& = &\fnd{\lambda}{2\mu_{c}a}J_{\ell }\left( k,a\right) A_{f}
\;\;\; , \\ [.3cm]
\fnd{1}{2\mu_{f}}\left(
\left[ J'_{\ell }\left( k,a\right)k^{2\ell +1}
\xrm{cotg}\left(\delta_{\ell }(E)\right)-
N'_{\ell }\left( k,a\right)\right] B_{f}-
J'_{\ell }\left( k,a\right) A_{f}\right) & = &
\fnd{\lambda}{2\mu_{c}a}F_{c,\ell}(E,a) A_{c}
\;\;\; .
\label{reqa1a}
\end{array}
\end{equation}

\noindent
By elimination of the normalization constants, one finds
for the cotangent of the phase shift the expression

\begin{equation}
k^{2\ell +1}\xrm{cotg}\left(\delta_{\ell }(E)\right)\; =\;
\fnd{N_{\ell }\left( k,a\right)}{J_{\ell }\left( k,a\right)}\; -\;
\left[\lambda^{2}\fnd{\mu_{f}}{\mu_{c}}\fnd{1}{a^{2}}
J^{2}_{\ell }\left( k,a\right)
\fnd{F_{c,\ell}(E,a)G_{c,\ell}(E,a)}{W\left(F_{c,\ell},G_{c,\ell}\right)}
\right]^{-1}
\;\;\; ,
\label{phshft1}
\end{equation}

\noindent
Using moreover the relations~(\ref{BN}), we arrive at

\begin{equation}
\xrm{cotg}\left(\delta_{\ell }(E)\right)\; =\;
\fnd{n_{\ell }\left( ka\right)}{j_{\ell }\left( ka\right)}\; -\;
\left[\lambda^{2}\fnd{\mu_{f}k}{\mu_{c}a^{2}}\;
j^{2}_{\ell }\left( ka\right)\;
\fnd{F_{c,\ell}(E,a)G_{c,\ell}(E,a)}{W\left(F_{c,\ell},G_{c,\ell}\right)}
\right]^{-1}
\;\;\; .
\label{phshft2}
\end{equation}

Note that, in Eq.~(\ref{phshft2}), we have obtained a simpler expression for
the phase shift than in Eq.~(\ref{partialpshift}), in particular numerically,
since the convergence of the sum in Eq.~(\ref{partialpshift}) is very slow.
It is, moreover, amusing that we have thus derived, as a side result, a
``simple'' (depending on one's taste) proof for the following
{\it spectral representation} \/of the Green's function:

\begin{equation}
\sum_{n=0}^{\infty}\fnd{\abs{{\cal F}_{n\ell}(a)}^{2}}
{E-E_{n\ell}}\; =\;\fnd{2\mu_{c}}{a^{4}}\;
\fnd{F_{c,\ell}(E,a)G_{c,\ell}(E,a)}
{W\left(F_{c,\ell}(E,a),G_{c,\ell}(E,a)\right)}
\;\;\; .
\label{Complitude}
\end{equation}

\noindent
This relation must be true for any confining potential with an infinite
discrete set of radial and angular excitations.
The set of functions $\left\{{\cal F}_{n\ell}\; ;\; n=0,1,2,\dots\right\}$
represents, for orbital angular momentum $\ell$,
a full set of radial eigensolutions, with eigenvalue $E_{n\ell}$,
of the Hamiltonian $H_{c}$.
Furthermore, $F_{c,\ell}$ and $G_{c,\ell}$ represent two linearly
independent solutions for any value of the energy $E$.

\section{More realistic ``phase'' transitions}
\label{Deltaell}

The results of Appendices~(\ref{Tmtrx}) and (\ref{Delta}) are illustrative
for studying the skeleton of the calculus involved in determining
the elastic scattering phase shifts,
but do not represent the coupling of quark-antiquark states to
meson-meson systems.
An easy way to understand this is by comparing the parities of both
phases of the system.
Due to Fermi statistics, the parity of the $q\bar{q}$ phase is given
by $P=(-1)^{\ell_{c}+1}$.
On the other hand, from Bose statistics follows that the parity of the
meson-meson phase is given by $P=(-1)^{\ell_{f}}$.
One concludes that the orbital angular momenta $\ell_{c}$ and $\ell_{f}$
for respectively the $q\bar{q}$ phase and the meson-meson phase,
must differ at least one unit.
Hence, the potential given in Eq.~(\ref{Opmodel}) cannot couple
the two different phases.
Nevertheless, expression (\ref{partialpshift}) for the scattering
phase shift is a powerful tool for further investigation.
In Appendix~(\ref{Delta}) we obtained a different form for this quantity.
Now the procedure leading from Eq.~(\ref{2x2weq2}) to
Eq.~(\ref{phshft2}) can be repeated for the more general case,
given by

\begin{eqnarray}
h_{c} & = & \fnd{1}{2\mu_{c}}
\left( -\fnd{d^{2}}{dr^{2}}\; +\;
\fnd{\ell_{c}\left(\ell_{c}+1\right)}{r^{2}}\right)\; +\;
m_{q}\; +\; m_{\bar{q}}\; +\; V_{c}(r)
\;\;\; ,
\nonumber \\ [.3cm]
h_{f} & = & \fnd{1}{2\mu_{f}}
\left( -\fnd{d^{2}}{dr^{2}}\; +\;
\fnd{\ell_{f}\left(\ell_{f}+1\right)}{r^{2}}\right)\; +\;
M_{1}\; +\; M_{2}
\;\;\; , \;\;\;\xrm{and}
\nonumber \\ [.3cm]
V & = & \fnd{1}{2\mu_{c}a}\;\delta\left( r-a\right)
\;\;\; .
\label{2x2f2dlt}
\end{eqnarray}

\noindent
Substituting $\ell$ in Eq.~(\ref{phshft2}) conveniently by either
$\ell_{c}$ or $\ell_{f}$,
one gets for the scattering phase shift the result

\begin{equation}
\xrm{cotg}\left(\delta_{\ell_{f}}(E)\right)\; =\;
\fnd{n_{\ell_{f}}\left( ka\right)}{j_{\ell_{f}}\left( ka\right)}\; -\;
\left[\lambda^{2}\fnd{\mu_{f}k}{\mu_{c}a^{2}}\;
j^{2}_{\ell_{f}}\left( ka\right)\;
\fnd{F_{c,\ell_{c}}(E,a)G_{c,\ell_{c}}(E,a)}
{W\left(F_{c,\ell_{c}},G_{c,\ell_{c}}\right)}
\right]^{-1}
\;\;\; .
\label{phshft3}
\end{equation}

\noindent
Then we can make use of Eq.~(\ref{Complitude}) to arrive at

\begin{equation}
\xrm{cotg}\left(\delta_{\ell_{f}}(E)\right)\; =\;
\fnd{n_{\ell_{f}}\left( ka\right)}{j_{\ell_{f}}\left( ka\right)}\; -\;
\left[\lambda^{2}\fnd{\mu_{f}ka^{2}}{2\mu_{c}^{2}}\;
j^{2}_{\ell_{f}}\left( ka\right)\;
\sum_{n=0}^{\infty}\fnd{\abs{{\cal F}_{n\ell_{c}}(a)}^{2}}
{E-E_{n\ell_{c}}}
\right]^{-1}
\;\;\; .
\label{phshft}
\end{equation}

\noindent
This is the result that can be used for our model.
However, note that we have not specified how we arrived from
Eq.~(\ref{c2x2weqCS}) to the relations (\ref{2x2f2dlt}).
Operators which provide for the communication between channels of different
orbital angular momenta and quark-antiquark spins, can be constructed
\cite{PRD27p1527,ZPC21p291,ZPC17p135},
but the procedure involves rather technical recoupling schemes of quantum
numbers.
We do not intend to go into the details here.

Extension of the formalism to potentials more realistic than a spherical
delta shell is also straightforward, and can be found in
Refs.~\cite{CPC27p377,LNP211p182}.
However, it is opportune to mention here that the solutions of the full model
\cite{ZPC30p615,PRD27p1527} do not behave differently from the simple
formula~(\ref{phshft}) in the energy domain under study in this work,
only yielding some minor differences in the numerical results.


\begin{thebibliography}{60}
\bibitem{Montanet2003}
L.~Montanet,
in {\it Nyiri, J. (ed.): The Gribov theory of quark confinement}, pp. 275-296.

\bibitem{NUCLEX0302007}
M.~Wolke {\it et al.},
arXiv:nucl-ex/0302007.

\bibitem{HEPPH0301126}
M.~B\"{u}scher, F.~P.~Sassen, N.~N.~Achasov and L.~Kondratyuk,
Contribution to the workshop on the future physics program at COSY-J\"{u}lich,
CSS2002,
arXiv:hep-ph/0301126.

\bibitem{HEPPH0302137}
V.~V.~Anisovich, L.~G.~Dakhno and V.~A.~Nikonov,
arXiv:hep-ph/0302137.

\bibitem{HEPPH0210431}
Abdou~M.~Abdel-Rehim, Deirdre~Black, Amir~H.~Fariborz, Joseph~Schechter,
Phys.\ Rev.\ D {\bf 67} (2003) 054001
[arXiv:hep-ph/0210431].

\bibitem{PLB559p49}
P.~Colangelo and F.~De Fazio,
Phys.\ Lett.\ B {\bf 559} (2003) 49
[arXiv:hep-ph/0301267].

\bibitem{HEPPH0212361}
Hai-Yang Cheng,
arXiv:hep-ph/0212361.

\bibitem{HEPPH0212117}
Hai-Yang~Cheng,
arXiv:hep-ph/0212117.

\bibitem{HEPPH0302059}
Chuan-Hung Chen,
arXiv:hep-ph/0302059.

\bibitem{HEPPH0303248}
S.~F.~Tuan,
arXiv:hep-ph/0303248.

\bibitem{HEPPH0304031}
A.~Faessler, T.~Gutsche, M.~A.~Ivanov, V.~E.~Lyubovitskij and P.~Wang,
arXiv:hep-ph/0304031.

\bibitem{HEPPH0303223}
D.~Black, M.~Harada and J.~Schechter,
arXiv:hep-ph/0303223.

\bibitem{PRD15p267}
Robert L.~Jaffe,
Phys.\ Rev.\ D {\bf 15} (1977) 267.

\bibitem{PRD66p010001}
K.~Hagiwara {\it et al.} \/[Particle Data Group Collaboration],
{\it Review Of Particle Physics},
Phys.\ Rev.\ D {\bf 66} (2002) 010001.

\bibitem{PRD26p239}
M.~D.~Scadron,
Phys.\ Rev.\ {\bf D26}, 239 (1982).

\bibitem{NPB320p1}
J.~E.~Augustin {\it et al.}  [DM2 Collaboration],
Nucl.\ Phys.\ B {\bf 320}, 1 (1989).

\bibitem{HEPEX0204018}
E.~M.~Aitala {\it et al.} [E791 Collaboration],
Phys.\ Rev.\ Lett.\  {\bf 89} (2002) 121801
[arXiv:hep-ex/0204018].

\bibitem{PRD59p074001}
J.~A.~Oller, E.~Oset and J.~R.~Pel\'{a}ez,
Phys.\ Rev.\ {\bf D59} (1999) 074001
[Erratum-ibid.\ {\bf D60} (1999) 099906]
[arXiv:hep-ph/9804209].

\bibitem{ZPC30p615}
E.~van Beveren, T.~A.~Rijken, K.~Metzger, C.~Dullemond, G.~Rupp and
J.~E.~Ribeiro,
Z.\ Phys.\ {\bf C30} (1986) 615.

\bibitem{EPJC22p493}
Eef van Beveren and George Rupp,
Eur.\ Phys.\ J.\ {\bf C22}, 493 (2001)
[arXiv:hep-ex/0106077].

\bibitem{PLB449p154}
A.~V.~Anisovich {\it et al.},
Phys.\ Lett.\ B {\bf 449} (1999) 154.

\bibitem{PLB517p261}
A.~V.~Anisovich {\it et al.},
Phys.\ Lett.\ B {\bf 517} (2001) 261.

\bibitem{PRD27p1527}
E.~van Beveren, G.~Rupp, T.~A.~Rijken, and C.~Dullemond,
Phys.\ Rev.\ {\bf D27} (1983) 1527.

\bibitem{PRD44p2803}
A.~G.~Verschuren, C.~Dullemond, and E.~van Beveren,
Phys.\ Rev.\ {\bf D44} (1991) 2803.

\bibitem{PLB541p22}
Claude Amsler,
Phys.\ Lett.\ B {\bf 541} (2002) 22
[arXiv:hep-ph/0206104].

\bibitem{HEPPH0301012}
L.~Ya.~Glozman,
arXiv:hep-ph/0301012.

\bibitem{HEPLAT0210012}
Teiji~Kunihiro, Shin~Muroya, Atsushi~Nakamura, Chiho~Nonaka,
Motoo~Sekiguchi and Hiroaki~Wada [SCALAR Collaboration],
Talk given at 20th International Symposium on Lattice Field Theory
(LATTICE 2002), Boston, Massachusetts, 24-29 Jun 2002,
arXiv:hep-lat/0210012.

\bibitem{HEPPH0302133}
Amir~H.~Fariborz,
arXiv:hep-ph/0302133.

\bibitem{PRL77p2332}
N.~Isgur and J.~Speth,
Phys.\ Rev.\ Lett.\  {\bf 77} (1996) 2332.

\bibitem{PLB361p160}
E. Klempt, B.C. Metsch, C.R. M\"{u}nz and H.R. Petry,
Phys.\ Lett.\ B {\bf 361} (1995) 160
[arXiv:hep-ph/9507449].

\bibitem{PRD63p014019}
C.~M.~Shakin and Huangsheng Wang,
Phys.\ Rev.\ D {\bf 63} (2001) 014019.

\bibitem{PRD65p114011}
C.~M.~Shakin,
Phys.\ Rev.\ D {\bf 65} (2002) 114011.

\bibitem{PRD65p078501}
George Rupp, Eef van Beveren and Michael D.~Scadron,
Phys.\ Rev.\ D {\bf 65} (2002) 078501
[arXiv:hep-ph/0104087].

\bibitem{EPJC10p469}
Eef van Beveren and George Rupp,
Eur.\ Phys.\ J.\ {\bf C10}, 469 (1999)
[arXiv:hep-ph/9806246].

\bibitem{CPC27p377}
C.~Dullemond, G.~Rupp, T.~A.~Rijken, and E.~van Beveren,
Comp.\ Phys.\ Comm.\ {\bf 27} (1982) 377.

\bibitem{PRD21p772}
E.~van Beveren, C.~Dullemond, and G.~Rupp,
Phys.\ Rev.\ {\bf D21} (1980) 772
[Erratum-ibid.\ {\bf D22} (1980) 787].

\bibitem{NC14p951}
Tullio~Regge,
Nuovo Cim.\  {\bf 14} (1959) 951.

\bibitem{PLB413p137}
A.~V.~Anisovich and A.~V.~Sarantsev,
Phys.\ Lett.\ {\bf B413} (1997) 137
[arXiv:hep-ph/9705401].

\bibitem{HEPPH0204328}
V.~V.~Anisovich and A.~V.~Sarantsev,
arXiv:hep-ph/0204328.

\bibitem{HEPPH0110156}
Eef van Beveren and George Rupp,
AIP Conf.\ Proc.\  {\bf 619} (2002) 209
[arXiv:hep-ph/0110156].

\bibitem{PRD60p034002}
Amir H.~Fariborz and Joseph Schechter,
Phys.\ Rev.\ {\bf D60}, 034002 (1999)
[hep-ph/9902238].

\bibitem{ZPC21p291}
E.~van Beveren,
Z.\ Phys.\ {\bf C21} (1984) 291.

\bibitem{EPJC11p717}
Eef van Beveren and George Rupp,
Eur.\ Phys.\ J.\ C {\bf 11} (1999) 717
[arXiv:hep-ph/9806248].

\bibitem{NPB587p331}
Matthias Jamin, Jos\'{e}~Antonio Oller, and Antonio Pich,
Nucl.\ Phys.\ {\bf B587}, 331 (2000)
[hep-ph/0006045].

\bibitem{PotentialScattering}
V. De Alfaro and T. Regge,
{\it Potential Scattering}, North Holland (Amsterdam, 1965).

\bibitem{HEPPH0203149}
M.~Boglione and M.~R.~Pennington,
Phys.\ Rev.\ D {\bf 65} (2002) 114010
[arXiv:hep-ph/0203149].

\bibitem{PLB521p15}
F.~De Fazio and M.~R.~Pennington,
Phys.\ Lett.\ B {\bf 521} (2001) 15
[arXiv:hep-ph/0104289].

\bibitem{NPB266p451}
R.~N.~Cahn and P.~V.~Landshoff,
Nucl.\ Phys.\ B {\bf 266} (1986) 451.

\bibitem{HEPPH0201006}
Eef van Beveren and George Rupp,
Proceedings of the Workshop on Recent Developments in Particle
and Nuclear Physics, April 30, 2001, Coimbra (Portugal) ISBN
972-95630-3-9, pages 1-16,
[arXiv:hep-ph/0201006].

\bibitem{NPA688p823}
S.~N.~Cherry and M.~R.~Pennington,
Nucl.\ Phys.\ A {\bf 688} (2001) 823
[arXiv:hep-ph/0005208].

\bibitem{PRD61p014015}
W.~Lee and D.~Weingarten,
Phys.\ Rev.\ {\bf D61}, 014015 (2000)
[hep-lat/9910008];
 
D.~Weingarten, private communication.

\bibitem{HEPLAT9805029}
W.~Lee and D.~Weingarten,
hep-lat/9805029.

\bibitem{NPPS53p236}
W.~Lee and D.~Weingarten,
Nucl.\ Phys.\ Proc.\ Suppl.\  {\bf 53}
[hep-lat/9608071].

\bibitem{ZPC68p647}
\author{Nils A. T\"{o}rnqvist}
Z.\ Phys.\ C {\bf 68} (1995) 647
[arXiv:hep-ph/9504372].

\bibitem{PRL76p1575}
Nils~A.~T\"{o}rnqvist and Matts~Roos,
Phys.\ Rev.\ Lett.\  {\bf 76} (1996) 1575
[arXiv:hep-ph/9511210].

\bibitem{PLB462p14}
Kim~Maltman,
Phys.\ Lett.\ {\bf B462} (1999) 14
[arXiv:hep-ph/9906267].

\bibitem{HEPPH0204205}
Frank~E.~Close and Nils~A.~T\"{o}rnqvist,
J.\ Phys.\ G {\bf 28} (2002) R249
[arXiv:hep-ph/0204205].

\bibitem{Bauer}
W. Bauer, Journal f\"{u}r Mathematik {\bf LVI} (1859), pp. 104-106;
see also:
G.N. Watson, {\it A treatise on the Theory of Bessel Functions},
section 4.32.

\bibitem{ZPC17p135}
E.~van Beveren,
Z.\ Phys.\ {\bf C17} (1983) 135.

\bibitem{LNP211p182}
E.~van Beveren, C.~Dullemond, T.~A.~Rijken, and G.~Rupp,
Lect.\ Notes Phys.\  {\bf 211} (1984) 182.
\end{thebibliography}
\end{document}